\documentstyle[aps,pre,floats,epsf]{revtex}
\begin{document}
\draft
\twocolumn[\hsize\textwidth\columnwidth\hsize\csname@twocolumnfalse%
\endcsname

\title{Smooth Phases, Roughening Transitions and Novel Exponents\\
in One-dimensional Growth Models.\\}
\author{U. Alon$^{1,2}$, M. R. Evans$^{3}$, H. Hinrichsen$^{4}$,
        and D. Mukamel$^{1}$\\[-2mm]$ $}
\address{
$^{1}$  Department of Physics of Complex Systems,
        Weizmann Institute, Rehovot 76100, Israel\\ 
$^{2}$  Present address: Depts. of Molecular Biology and Physics,
        Princeton University, Princeton NJ 08540\\
$^{3}$  Department of Physics and Astronomy, University of Edinburgh,
        Mayfield Road, Edinburgh EH9 3JZ, U.K.\\
$^{4}$  Max-Planck-Institut f\"ur Physik komplexer Systeme,
        N{\"o}thnitzer Stra{\ss}e 38, D-01187 Dresden, Germany\\
        }
\date{October 14, 1997, submitted to Phys. Rev. E}
\maketitle
\begin{abstract}
A class of solid-on-solid  growth models with short range
interactions and sequential updates is studied. The models exhibit
both smooth and rough phases in dimension $d=1$.  Some of the features
of the roughening transition which takes place in these models are
related to contact processes or directed percolation type
problems. The models are analyzed using a mean field approximation,
scaling arguments and numerical simulations. In the smooth phase the
symmetry of the underlying dynamics is spontaneously broken.  A family
of order parameters which are not conserved by the dynamics is defined
as well as conjugate fields which couple to these order
parameters. The corresponding critical behavior is studied and
novel exponents identified and measured. We also show how continuous
symmetries can be broken in one dimension.  A field theory appropriate
for
studying the roughening transition is introduced and discussed.

\end{abstract}
\pacs{PACS numbers: \ 05.40.+j; 05.70.Ln; 68.35.Fx; 64.60.Ak }]
%
%
%
%
%
%
\section{Introduction}
\label{IntroductionSection}
%
%
The statistical properties of moving interfaces and
surfaces of growing crystals
have been studied extensively in recent years~\cite{GeneralGrowth}. Various 
theoretical approaches have been applied in these studies including
field theoretical analyses of continuum KPZ-type 
equations~\cite{KPZ,KrugSpohn,Halpin} and
studies of discrete growth processes such as solid-on-solid (SOS) or
polynuclear growth (PNG) models amongst others 
\cite{Halpin,Zia,Rich,Goldenfeld,KK,KW,KKW,RSLP,TFW,GW,NdN}.

One of the more interesting properties of moving interfaces is their
roughness.  It is well known that in $d > 2$ dimensions, moving
interfaces may be either rough or smooth depending on the level of
the noise in the system.  However, growth processes in $d=1$ dimensions
are more subtle. Most growth processes governed by short range
interactions, such as those described by the KPZ equation yield a
rough interface. On the other hand study of a class of PNG models
suggested that both smooth and rough phases may exist in one dimension
($1d$)~\cite{KW,LRWK,Toom}.  However, this class of models is
characterized by two key features: (a) The evolution takes place by a
parallel updating process in which time progresses in discrete steps
and all sites of the interface are updated simultaneously according to
a given rule at any given time step. Such dynamics tend to be less
noisy than sequential updating processes, in which one site is updated
at a time. (b) The models have a maximal velocity by which the
uppermost point of the surface can propagate.  The existence of a
maximal velocity in these models is due to the use of parallel
updates, and the smooth phase disappears when random sequential
(continuous time) updates are used. An interesting question is whether a
sequential update growth process is capable of exhibiting smooth and
rough phases in $1d$.

Some time ago a transition from a smooth to a rough surface was 
observed in a SOS model
for surface reconstruction with sequential updates and particle 
conservation~\cite{RSLP}. The free parameter, $H$, is the
maximal allowed height difference between adjacent sites and is a
discrete quantity. It was observed that the surface is smooth 
for $H<3$ and macroscopically
grooved for $H>3$. At $H=3$  the surface appears to be rough. 
This phenomenon is clearly related to the local conservation law 
in this model.

Recently, a class of growth processes with short range interactions,
sequential updates and non-conserved dynamics
was introduced~\cite{Us}. They were
demonstrated to exhibit both rough and smooth phases in $1d$. The
dynamics is described by SOS models 
in which adsorption and desorption
processes take place in a ring geometry.  Depending on the relative
rates of the two processes one finds either a smooth or a rough
phase. In studying the roughening transition in these models, it has
been found that some of its features are related to those of contact
processes, or directed percolation, which have been extensively
studied in the past~\cite{DP,Durrett,CS,DK}.  These latter systems exhibit a
continuous phase transition which is strongly linked to the existence
of absorbing states (a set of states from which the system can not
escape). The model introduced in~\cite{Us} may be viewed as composed
of a series of contact processes interacting with each other, whose
dynamics does not have an absorbing state. This model is the subject
of the present paper.

An intriguing question related to the existence of a smooth phase is
that of spontaneous symmetry breaking (SSB) and long-range order in 1d
systems with short-range interactions and small but unbounded noise.
By unbounded noise we mean that in a finite system all microscopic
configurations can be reached from any initial condition in a finite
time.  It is well known that in thermal equilibrium no phase
transition takes place under these conditions. Systems far from
equilibrium~\cite{Note} such as moving interfaces or driven diffusive
systems are, however, less restrictive and the question of
existence of SSB under these conditions has been open for some time.
Recently, a simple non-equilibrium model which exhibits SSB in $1d$
was introduced~\cite{SSBConserved1,SSBConserved2}.  This model belongs
to a class of driven-diffusive systems, in which charges of two kinds
are injected at the ends of a 1d lattice and are biased to move in
opposite directions. The dynamic rules are symmetric under space and
charge inversion (PC invariance). However, this symmetry is broken in
the stationary state of the system where the currents of the two
charge species are different. In this model, SSB is the result of the
{\em conserved} order parameter in the bulk (charges are not created
or annihilated except at the boundaries), and the existence of {\em
open} boundary conditions. These two conditions favor the emergence of
SSB; the conserved dynamics slows down the temporal evolution of the
system, moreover flips from one broken symmetry phase to another can
originate only at the boundaries where the order parameter is not
conserved. Initial attempts to modify this model such that either or
both of these features are eliminated resulted in symmetric stationary
states with no SSB.

The growth model discussed in the present work provides a simple
example for SSB which takes place far from equilibrium. 
The breaking of symmetry takes place in the smooth phase and the
order parameter associated with this transition is not conserved by
the dynamics. The model thus demonstrates that SSB can take place in
$1d$ in non-equilibrium systems with non-conserved 
order parameter and ring geometry.

In this paper we present a detailed analysis of the growth models
introduced in~\cite{Us}. The relation of the models to contact
processes and directed percolation is discussed and several families
of novel critical exponents are defined.  One family describes the
critical behavior of magnetization-like order parameters related to
the symmetry breaking. Another family is related to statistical
properties of the interface height near the roughening transition.
The dynamical equations are analyzed using the mean field
approximation, and a field theoretical model appropriate for the
roughening transition is introduced.  The scaling picture that emerges
is far from complete but points to the existence of complex critical
behavior.

The paper is organized as follows: The growth model is defined in
Sec.~\ref{DefinitionSection}.  The relation to contact processes and
directed percolation is discussed in Sec.~\ref{DPSection}. In
Sec.~\ref{NumericalSection} the results of scaling analysis and
numerical studies are presented.  The question of spontaneous symmetry
breaking, and the family of order parameters and their critical
behavior are discussed in Sec.~\ref{SSBSection}. The dynamics of the
model is studied within the mean field approximation in
Sec.~\ref{MeanFieldSection}, and a field theoretical model appropriate
for studying the critical behavior of the roughening transition is
defined and discussed in Sec.~\ref{FieldTheorySection}.  In
Sec.~\ref{PNGSection} some light is shed on the relation of the model
to the polynuclear growth models discussed above and in
Sec.~\ref{Conclusions} the main results are summarized and conclusions
drawn.  Finally, a particular case for which steady state can be
calculated exactly is presented in Appendix~\ref{SolvableCaseSection}.
%
%
%
\section{Model Definitions}
\label{DefinitionSection}
%
%
The class of models is most simply introduced in terms of the growth
of a one dimensional interface, in which both adsorption and
desorption processes take place. In the present models the key feature
is that desorption may take place only at the edge of a plateau,
i.e. at a site which has at least one neighbor at a lower height.  We
study two particular models in this class, (a) a restricted solid on
solid (RSOS) version where the height differences between neighboring
sites are restricted to take values $1,0,-1$ and (b) an unrestricted
model where there is no such restriction.  Both models are defined on
a 1d lattice of~$N$ sites $i=1\ldots N$ and associated with each site
is an integer height variable~$h_i$ which may take values $0,1 \ldots
\infty$.  Periodic boundary conditions are imposed so that $h_{N+i} =
h_{i} $.

The dynamics are defined through the following algorithm: at each
update choose a site~$i$ at random and carry out one of the following
processes: Adsorption of an adatom
\begin{equation}
\label{dyn1}
h_i \rightarrow h_{i}+1 \; \; \mbox {with probability } q 
\end{equation}
or desorption of an adatom from the edge of a step
\begin{eqnarray}
\label{dyn2}
h_i \rightarrow \mbox{min}(h_{i},h_{i+1}) \; \; 
\mbox {with probability } (1-q)/2\\
\label{dyn3}
h_i \rightarrow \mbox{min}(h_{i},h_{i-1}) \; \; 
\mbox {with probability } (1-q)/2
\end{eqnarray}
The update of the chosen site~$i$ is conveniently implemented in a
simulation by drawing a random number between 0 and 1 from a flat
distribution. If the number is less than~$q$ process (\ref{dyn1}) is
executed, if the number is greater than $(1+q)/2$ process (\ref{dyn2})
is executed, otherwise process (\ref{dyn3}) is executed.

In the RSOS version, a process is only carried out if it respects the
constraint
\begin{equation}
|h_{i}-h_{i+1}| \leq 1 \;.
\label{RSOS}
\end{equation}

For both models, when the growth rate~$q$ is low, the desorption
processes (\ref{dyn2})--(\ref{dyn3}) dominate. If all the heights are
initially set to the same value this layer will remain the bottom
layer of the interface. Small islands will grow on top of the bottom
layer through the process (\ref{dyn1}) but will quickly be eliminated
by desorption at the island edges.  Thus, the interface is effectively
anchored to its bottom layer and the velocity, defined as the rate of
increase of the minimum height of the interface, is zero in the
thermodynamic limit. On a finite system, very large fluctuations will
occasionally occur which allow a new layer to be completed and the
velocity will be positive but exponentially small in the system size.

As $q$ is increased the production of islands on top of the bottom
layer increases until above~$q_c$, the critical value of~$q$, the
islands merge and new layers are formed at a finite rate giving rise
to a finite interface velocity in the thermodynamic limit. 
Thus, above the transition one expects the velocity to behave as
\begin{equation}
v \sim (q-q_c)^y\,,
\label{ydef}
\end{equation}
where $y$ is  the critical exponent describing the growth
in velocity.

%
%
\begin{figure}
\epsfxsize=90mm         
\centerline{ \epsffile{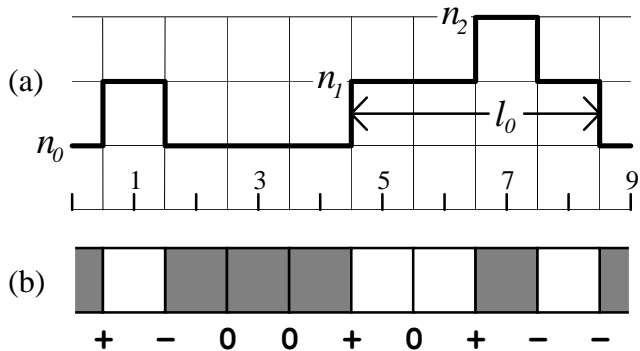}  } 
\caption {
(a) Typical configuration of the interface. $n_k$ is the fraction of
   sites at height~$k$ above the minimal height in the configuration
   (here $n_0=n_1=\frac49$, $n_2=\frac19$). The average island size grown 
   on top of level $k$ is $l_k$. 
(b) Mapping of the configuration of Fig.1a to the charged-particle 
   representation (see Eq.~\ref{ChargedRep}), 
   along with a site coloring, as described 
   in Sec.~\ref{SSBSection}.}
\label{Deffig}
\end{figure}

Another critical exponent is defined by considering~$n_0$, the
fraction of sites at the lowest level (see Fig.~\ref{Deffig}). 
Below the transition $(q<q_c)$
this fraction will be large since the interface is anchored at this
level. As the transition is approached more and more islands form on
top of the bottom layer and the fraction $n_0$ will decrease until it
vanishes at the transition. This may be described by
\begin{equation}
n_0 \sim (q_c-q)^{x_0}\;.
\label{x0def}
\end{equation}
Similarly one may define a family of exponents $\{ x_k \}$ describing
the vanishing of $n_k$, the fraction of sites at level~$k$,
\begin{equation}
n_k \sim (q_c-q)^{x_k}\;.
\label{xkdef}
\end{equation}
The interface width is defined by
\begin{equation}
W=[N^{-1} \sum_i (h_i - N^{-1}\sum_j h_j)^2]^{1/2}\; .
\label{width}
\end{equation}
Below the transition the width should be finite indicating a smooth
interface exploring only a finite height above the bottom layer.
However, above the transition the interface should display the behavior
generic to moving interfaces~\cite{KPZ}, that is roughening where the
width diverges as
\begin{equation}
W \sim N^{1/2} \qquad \mbox{at} \ q>q_c\,.
\end{equation}
Therefore near to and above the transition we expect
\begin{equation}
W \sim N^{1/2} (q-q_c)^{\chi}\;,
\label{chidef}
\end{equation}
where $\chi$ is the exponent describing  the vanishing
of the roughness as the transition is approached.

The above considerations hold for both RSOS and unrestricted models and 
 and  we will address the question as to what extent the two
models can be seen as representatives of a whole  class
of models with the same critical behavior.  In 
Sec.~\ref{NumericalSection} we will further explore the scaling
behavior, adding to the exponents $y, x_k, \chi$ that we have so far
introduced.  However, we defer this until after 
Sec.~\ref{DPSection} where we discus 
the relation to a directed percolation
model through which some of the simple critical behavior may be
understood.

For the moment we note that the RSOS version (\ref{RSOS}) may be
viewed as a driven diffusion model of two oppositely charged types of
particles. The charges
\begin{equation}
\label{ChargedRep}
c_{i} \;=\; h_{i}-h_{i-1} \;\in\; \{-\ ,\; 0\ ,\;+\}
\end{equation}
are bond variables and
represent a change of height between adjacent interface sites 
(see Fig.~\ref{Deffig}).
In this representation, 
it is convenient to convert
the dynamical rules (\ref{dyn1})-(\ref{dyn3})
into processes occurring at bonds with the following {\it rates}
\begin{eqnarray}
\mbox{0\ +} \;\; &\rightarrow &\;\; \mbox{+\ 0} \;\;\;\mbox{with
rate}\;\; g
\nonumber \\
\mbox{+\ 0} \;\; &\rightarrow  &\;\; \mbox{0\ +} \;\;\;\mbox{with
rate}\;\; 1
\nonumber \\
\mbox{$-$\ 0} \;\; &\rightarrow &\;\; \mbox{0\ $-$} \;\;\;\mbox{with
rate}\;\;
g
\nonumber \\
\mbox{0\ $-$} \;\; &\rightarrow  &\;\; \mbox{$-$\ 0} \;\;\;\mbox{with
rate}\;\;
1
\nonumber \\
\mbox{0\ \ 0} \;\; &\rightarrow &\;\; \mbox{+\ $-$} \;\;\mbox{with
rate}\;\; g
\nonumber \\
\mbox{+\ $-$} \;\; &\rightarrow  &\;\; \mbox{0\ \ 0} \;\;\;\mbox{with
rate}\;\;
2
\nonumber \\
\mbox{$-$\ +} \;\; &\rightarrow  &\;\;\mbox{0\ \ 0} \;\;\;\mbox{with
rate}\;\;
g
\end{eqnarray}
where the growth rate $g$ is related to $q$ of
equations (\ref{dyn1})--(\ref{dyn3}) through
\begin{equation}
g = 2q/(1-q) \, .
\label{gdef}
\end{equation}
The rule that desorption cannot occur from the middle of a plateau
corresponds to the absence of the process $\mbox{0\ 0} \rightarrow
\mbox{$-$\ +} $.  In this version the dynamics may be performed
without reference to the height variables $h_i$ (although these can
easily be reconstructed to within the height of the bottom layer from
the variables $c_{i}$).  The critical behavior is reflected in the
correlations between the charges.  At $q<q_c$, the charges are
arranged as closely bound $+\ -$ dipoles.  At $q>q_c$, the dipoles
become unbound, and the fluctuations in the total charge, measured
over a distance of order $N$, diverge with $N$.

The charged particle representation also allows the effect of
desorption from the middle of a plateau to be studied. This is done in
Appendix A by introducing a non-zero rate $p$ for the process $0\, 0
\rightarrow  -\ +$ and solving the steady state exactly in the case
$p=1-g/2$. The result is that, although different choices of this rate
allow the interface velocity to be positive, zero or negative, the
interface is always rough and no smooth phase exists. 
%
%
%
%
\section{Relation to Directed Percolation and Contact Processes}
\label{DPSection}
%
%
Some of the critical behavior described in the previous section may
be related to that of directed percolation (DP).  In DP sites of a
lattice are either occupied by a particle or empty.  The dynamical
processes are that a particle can self-annihilate or produce an
offspring at a neighboring empty site. If the rate of self-annihilation
is sufficiently high the system always reaches an absorbing state where
no particles remain and therefore no further particle can be produced. 
On the other hand when the rate of offspring production is high,
another steady state of the system, where the density of particles is
finite, exists on the infinite lattice and is termed the active phase.

In DP the dynamics is usually carried out in parallel,
e.g.~\cite{DK}. In the corresponding models in the mathematical
literature, known as contact processes~\cite{Durrett}, the dynamics
are defined in continuous time which can be numerically implemented by
random sequential dynamics.

To see the analogy with the growth model defined in 
Sec.~\ref{DefinitionSection} consider the dynamics of the bottom layer of
the unrestricted model. Let us introduce variables $\{ s_i \}$ which
take value 1 if the height of site $i$ is that of the bottom layer and
take value zero otherwise. The algorithm stated in
(\ref{dyn1})--(\ref{dyn3}) is exactly equivalent to the following
dynamics for the $\{ s_i \}$.  At each update randomly select a site
$i$ and modify $s_i$ with the following probabilities
\begin{eqnarray}
\label{DP}
\lefteqn{\mbox{if}\;\; s_i = 1}\nonumber \\
&&\hspace{0.5in} s_i \rightarrow 0 \; \; \mbox {with prob. } q
\nonumber\\
\lefteqn{\mbox{if}\;\;\{ s_{i-1}, s_i, s_{i+1}\} = \{ 0, 0, 1 \}}
\nonumber \\
&&\hspace{0.5in}s_i \rightarrow 1 \; \; \mbox {with prob. } (1-q)/2 \\
\lefteqn{\mbox{if}\;\;\{ s_{i-1}, s_i, s_{i+1}\} = \{ 1, 0, 0 \}}
\nonumber \\
&&\hspace{0.5in}s_i \rightarrow 1 \; \; \mbox {with prob.
}(1-q)/2\nonumber\\
\lefteqn{\mbox{if}\;\;\{ s_{i-1}, s_i, s_{i+1}\} = \{ 1, 0, 1 \}}
\nonumber \\
&&\hspace{0.5in} s_i \rightarrow 1  \; \; \mbox {with prob. } 1-q 
\nonumber
\end{eqnarray}
This dynamics is exactly that of a contact process when we take $s_i
=1$ to indicate the occupation of a site: the particles
self-annihilate with rate $\lambda = q/(1-q)$ and a particle is
created at a vacant site with rate 1/2 if one neighbor is occupied and
rate 1 if both neighbors are occupied. This process has been
extensively studied by series expansions~\cite{Dickman} and short time
expansions~\cite{JD} and the transition found to occur at $\lambda_c\simeq
0.3032$ corresponding to $q_c \simeq 0.2327$ for the unrestricted growth
model.  Thus, as seen by the bottom layer of the growing interface, the
transition from anchored to moving phase is simply a DP transition.
The anchored phase corresponds to the active DP phase whereas the
moving phase corresponds to the absorbing DP phase.

The critical behavior of DP may be described as follows.  
Above the transition ($\lambda>\lambda_c$)
an initial seed particle will typically produce activity
over a region of lateral extent $\xi_{\perp} \sim
|\epsilon|^{-\nu_{\perp}}$ and duration $\xi_{\parallel} \sim
|\epsilon|^{-\nu_{\parallel}}$, before the absorbing state is reached.
Here $\epsilon$, given by
\begin{equation}
\epsilon = q-q_c\; ,
\label{eps}
\end{equation}
measures how far the system is from criticality and $\xi_{\perp}$ and
$\xi_{\parallel}$ are interpreted as spatial and temporal scaling
lengths which diverge at the transition.  
Below the transition ($\lambda<\lambda_c$) the
density $n$ of occupied sites in the active phase is $n \sim |
\epsilon|^{\beta}$ and there is a finite probability $|
\epsilon|^{\beta}$ that an initial seed particle will result in the
active phase being reached.  The lateral extent of such an active
cluster grows with time as $t^z$ where $z=\nu_\parallel/\nu_\perp$ is
the dynamic exponent.  The typical size $l$ of regions containing no
activity diverges as the transition is approached as $l \sim n^{-1}
\sim | \epsilon|^{-\beta}$.

These exponents allow one to readily identify $y$ and~$x_0$ defined in
Eqs. (\ref{ydef}) and (\ref{x0def}).  Since $n_0$ corresponds to the
density of occupied sites in the DP active phase we expect in
(\ref{x0def})
\begin{equation}
n_0 \sim (q_c-q)^{\beta} \;\;\mbox{and}\;\; x_0 = \beta \, .
\label{n0scal}
\end{equation}
In the moving phase the velocity is given by the inverse
of the typical time for the bottom layer to be covered over.
We identify this time scale with the lifetime of active
regions in the DP absorbing phase and we expect in (\ref{ydef})
\begin{equation}
v \sim 1/ \xi_\parallel \sim (q- q_c)^{\nu_\parallel}\;\;
\mbox{and}\;\;y=\nu_\parallel\,.
\label{velscal}
\end{equation}
Thus we see that the two exponents $x_0$ and $y$ that describe
quantities involving {\it only} the dynamics of the bottom layer may
be directly identified with known (numerically) DP exponents.
Exponents describing quantities involving higher levels of the growth
process such as $x_k$ given by (\ref{xkdef}) are not so
straightforward as we shall see in Secs.~\ref{NumericalSection} and
\ref{SSBSection}.

In this section the exact mapping of the bottom layer of the
unrestricted growth model to DP has been described.  For the RSOS model
there is no such exact mapping, nevertheless we expect the bottom layer
to exhibit DP behavior and relations (\ref{n0scal}),(\ref{velscal}) to
hold, because the phase transition  in this model should display a
robustness with respect to the microscopic rules similar
to that found in DP models.
%
%
\section{Scaling and numerical results} \label{NumericalSection}
%
%
%
So far we have seen that as criticality is approached from the smooth
phase the scaling properties of some quantities involving only the
bottom level of the interface may be adequately described using DP
exponents.  However, for more general quantities the scaling is less
clear cut and indeed only a partial picture emerges.  We first deal
with properties of the first few layers in the smooth phase using
heuristic arguments and then comparing 
them to numerical results. The width $W$, a
quantity involving all layers, is studied next. It is argued that a
simple scaling picture, involving only the DP correlation length
$\xi_{\perp}$, does not hold.  We provide evidence to suggest that
other length scales may be important as criticality is approached from
within the moving phase.  A partial scaling picture which emerges is
then summarized.

\subsection{Scaling properties of the first few layers}
\label{NumSubsectionA}
We now discuss the scaling properties of the first few
layers $k=1,2,\ldots$ above the bottom layer. Since,
in the smooth phase, the scaling
properties at the bottom layer ($k=0$) are completely characterized by the
three DP exponents $x_0=\beta$, $\nu_\perp$, and $\nu_{||}$, it is
natural to assume that the next layers obey similar scaling laws with
analogous exponents, $x_k=\beta^{(k)}$, $\nu_\perp^{(k)}$, and
$\nu_{||}^{(k)}$, where, for example, $x_k$ is the density exponent
defined in Eq.~(\ref{xkdef}). In principle all these exponents could
be different and independent. Our numerical results, however, suggest
that the scaling exponents $\nu_\perp^{(k)}$ and $\nu_{||}^{(k)}$ are
{\em identical on all levels} and equal to the DP exponents
$\nu_\perp$ and $\nu_{||}$. This remarkable property implies that the
growth process at criticality is characterized by a {\em single}
anisotropy exponent $z=\nu_{||}/\nu_\perp$. On the other hand we find
numerically that the density exponents $x_k$ for $k=1,2,\ldots$ are
different and considerably reduced compared to their DP value
$x_0=\beta$.  It appears that these exponents are non-trivial in
the sense that they are not simply related to DP exponents.

In order to explain the reduced values of $x_k$, we present a simple
heuristic argument~\cite{Us}. Let us first consider the bottom layer
$k=0$.  As explained in the previous section, DP is characterized by
two length scales: the average size of inactive islands $l_0 \sim
n_0^{-1} \sim |\epsilon| ^{-\beta}$, and the spatial scaling length
$\xi_{\perp}\sim |\epsilon|^{-\nu_{\perp}}$~\cite{kinzel}.  They both
are related, for a system of size $N$, by the finite size scaling
relation $l_0 \sim |\epsilon|^{-\beta} f(N |\epsilon|^{\nu_{\perp}})$,
where $f$ is a function satisfying $f(s) \sim s^{\beta/\nu_{\perp}}$
for $s \rightarrow 0$, implying that on a finite system at criticality
one has $n_0 \sim \l_0^{-1} \sim N^{-\beta/\nu_{\perp}}$. Now consider
the next layer $k=1$. One may view the sites at heights $k \ge 2$ as
islands of active sites growing on top of the inactive islands of the
$k=1$ level, whose typical size is $l_0$.  Applying the same scaling
relations and assuming that the system size may be replaced by $l_0$,
we find $n_1 \sim \l_1^{-1} \sim l_0^{-\beta/\nu_{\perp}}$, where
$l_1$ is the mean size of islands of sites with height $k \ge 2$.
Repeating this argument one obtains $n_k \sim l_k^{-1} \sim
\l_{k-1}^{-\beta/\nu_{\perp}}$ and therefore
\begin{equation}
\label {eq nk}
n_k \sim |\epsilon|^{x_k}, \;\;\;\;  x_k=\beta (\beta/\nu_{\perp})^k\,.
\end{equation}
Inserting the numerically known DP exponents~\cite{JensenDPExponents}
%
%
$$
\label{DPValues}
\beta=0.27649(4), \hspace{5mm}
\nu_{||}=1.73383(2) , \hspace{5mm}
\nu_{\perp}=1.09684(2)
$$
%
%
we obtain the approximations $x_1 \simeq 0.070$, $x_2 \simeq 0.017$,
and $x_3 \simeq 0.004$ which are in qualitative agreement with the
numerical results (see below).  However, this simple scaling argument is 
not quantitatively correct for several reasons.  First we consider the
temporal average of the island size $l_0$ as a fixed `finite size'
length for a DP process taking place on top of the island; we thus
neglect the temporal fluctuations of $l_0$.  
In addition, unlike exact scaling relations, our scaling argument
is expected to fail in higher dimensions since it is derived 
in the case of one dimension where active sites separate inactive island.
Nevertheless it can be used as a rough approximation as well as
qualitative explanation for the strongly reduced values of $x_k$
compared to $x_0$ for $k \geq 1$.

\subsection{The first few layers - numerical results}
%
%
\begin{figure}
\epsfxsize=90mm         
\epsffile{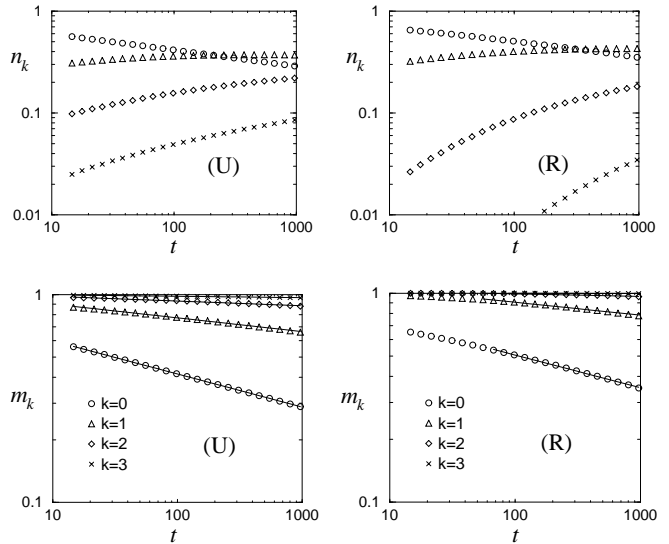}
\caption{
Dynamic simulations: the densities $n_k$ and the integrated densities
$m_k$ at criticality as a function of time $t$ in the
(U) unrestricted and (R) restricted growth model.  See
Eq.~(\ref{Method1}). }
\label{FigDynamicMC}
\end{figure}
In order to determine the density exponents $x_k$ and to verify that
the scaling exponents $\nu_\perp^{(k)}$ and $\nu_{||}^{(k)}$ are
indeed identical, we employ three different variants of Monte-Carlo
(MC) simulations, termed dynamic, static and finite size simulations,
described as follows.\\[2mm] 
\noindent {\it Dynamic simulations}: First we measure the temporal
evolution of the densities $n_k$ at criticality in a large system,
starting from a flat interface and averaging over a large number of
runs. For times larger than some transient time
the densities are expected to decay according to
\begin{equation}
\label{Method1}
n_k \sim t^{-x_k/\nu_{||}}\,.
\end{equation}
\noindent {\it Static simulations}: We also determine $x_k$ directly
in off-critical (static) simulations, measuring the densities $n_k$ in
a sufficiently large system in the smooth phase and averaging over
very long times.  Using this method we can determine the exponents
$x_k$ directly through the expected behavior
\begin{equation}
\label{Method2}
n_k \sim |\epsilon|^{x_k}\,,
\end{equation}
where $\epsilon=q-q_c<0$.\\[2mm]
\noindent{\it Finite size simulations}:
Finally we analyze the finite size scaling of $n_k$ 
in critical systems of size $N$ averaged over long times.
Here the expected scaling behavior reads
\begin{equation}
\label{Method3}
n_k \sim N^{-x_k/\nu_{\perp}}\,.
\end{equation}
%
%

%
%
%
\begin{figure}
\epsfxsize=90mm         
\epsffile{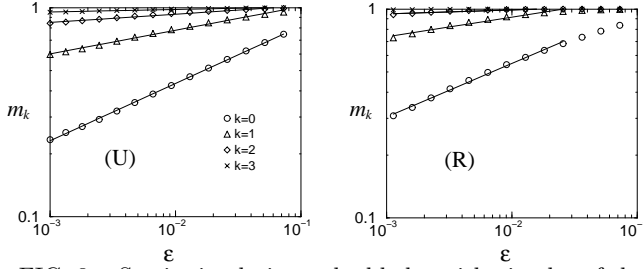}
\caption{
Static simulations: double logarithmic plot
of the saturation value of the integrated densities $m_k$
vs $|\epsilon|$. See Eq.~(\ref{Method2}). }
\label{FigStaticMC}
\end{figure}
%
%
%
%
\begin{figure}
\epsfxsize=90mm         
\epsffile{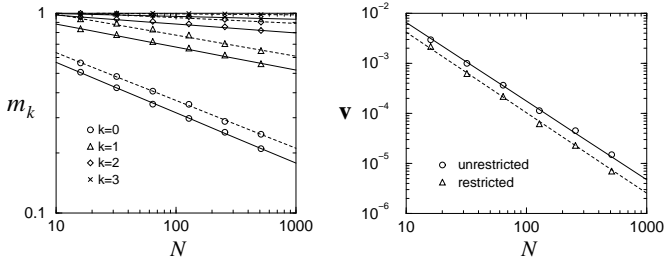}
\caption{
Finite size simulations: Measurement of the integrated densities $m_k$
and the growth velocity~$v$ at criticality as function of the system
size $N$. The solid (dashed) lines refer to
the unrestricted (restricted) model. See Eq.~(\ref{Method3}).}
\label{FigFiniteSizeMC}
\end{figure}

Thus the dynamic simulation should yield a numerical value for
$x_k/\nu_{||}$, the static simulation a value for $x_k$ and the finite
size simulation a value for $x_k/\nu_\perp$.  If, on inserting the DP
values of $\nu_{||}$ and $\nu_{\perp}$, the three methods
(\ref{Method1})-(\ref{Method3}) lead to the same numerical result for
the exponent $x_k$, to within numerical errors, we may conclude that
the scaling exponents $\nu_\perp^{(k)}$ and $\nu_{||}^{(k)}$ are
indeed equal to the DP exponents.\\

We observed that the quality of the numerical results can be improved
considerably if we measure the {\em integrated densities}
\begin{equation}
\label{IntegratedDensities}
m_k = \sum_{j=0}^k n_j
\end{equation}
instead of the densities $n_k$. The integrated density $m_k$ is the
probability of finding the interface at height $h \leq k$.  Since
$x_{k-1}>x_k$, we have $n_{k-1} \ll n_k$ near criticality and
therefore $m_k$ and $n_k$ scale asymptotically with the same
exponents. The difference between the two quantities is illustrated
for the case of dynamic simulations at $q_c$ in Fig.~\ref{FigDynamicMC}
(for the method of determining $q_c$ see below). As one
can see, the graph for the integrated densities $m_k$ in the
unrestricted model shows almost perfect straight lines in a double
logarithmic representation while the corresponding curves for $n_1$,
$n_2$ and $n_3$ still increase which makes it impossible to determine
the corresponding exponents with the $1000$ time steps
illustrated.  The same observation, although with less numerical
accuracy, holds for the restricted model. Therefore, instead of $n_k$,
we always measure the integrated densities $m_k$ for which we assume
the same scaling as in Eqs.~(\ref{Method1})-(\ref{Method3}).

\begin{table}
\begin{tabular}{|c||c|c|c|c|}
        && dynamic method      & static method  & finite size method
        \\\hline
$x_0$   & U: & $0.275(5)$      & $0.273(10)$    & $0.276(5)$\\
        & R: & $0.270(10)$     & $0.277(10)$    & $0.265(10)$\\\hline
$x_1$   & U: & $0.114(5)$      & $0.110(10)$    & $0.125(5)$\\
        & R: & $0.108(10)$     & $0.110(10)$    & $0.118(10)$\\\hline
$x_2$   & U: & $0.039(15)$     & $0.035(15)$    & $0.045(10)$\\
        & R: & $0.022(15)$     & $0.025(20)$    & $0.033(15)$\\\hline
$x_3$   & U: & $0.011(10)$     & no result      & $0.015(10)$\\
        & R: & no result       & no result      & no result\\
\end{tabular}
\vspace{2mm}
\caption{
Numerical estimates for the density exponents $x_0,\ldots,x_3$ for (U)
 the unrestricted and (R) the restricted version of the growth model
 obtained by various simulation methods.}
\end{table}

The dynamic MC simulations are performed on a large system with
$N=10^4$ sites starting from a flat interface. Detecting deviations
from the power law behavior, we find the critical points
$q_c=0.23267(3)$ for the unrestricted and $q_c=0.1889(1)$ for the
restricted model (corresponding to $g=0.4658$ in Eq. (\ref{gdef})).  
The time dependence of $m_k$ at criticality is
averaged over typically $10^5$ independent runs. The results are shown
in Fig.~\ref{FigDynamicMC}. From the slopes we estimate the critical
exponents $x_k/\nu_{||}$. Using $\nu_{\parallel}$
of DP we obtain the exponents $x_k$ which are summarized
in the left column of Table~I.

We note a numerical puzzle we have so far failed to explain.
The critical value of $q$ for the restricted model appears to be given
to high numerical accuracy by $q_{\rm u}/(1- q_{\rm u})$ where  $q_{\rm
u}$ is the  critical value of the unrestricted model. Whether this is
sheer coincidence or whether it is a manifestation of some duality
between the two models is an open question.

Static simulations are then carried out in the smooth phase $q<q_c$.
Varying $| \epsilon |=q_c-q$ from $10^{-3}$ to $10^{-1}$ we first
equilibrate the interface on a large lattice with $N=10^4$ sites
over a time interval up to $10^5$ time steps. Then the stationary
densities $m_k$ are averaged over a time interval of the
same size. The results are again averaged over $100$ independent
runs. Using this method we can estimate the critical exponents
$x_k$ directly from the slopes of the lines in Fig.~\ref{FigStaticMC}
(see middle column in Table~I). 

Finally finite-size simulations at criticality are carried out for
various lattice sizes $N=8,16,32,\ldots,1024$. Starting from a flat
interface we averaged the integrated densities $m_k$ over time
intervals proportional to $N^z$, ranging from $5000$ time steps for
$N=8$ up to $3 \cdot 10^7$ time steps for $N=1024$. Since finite sized
systems at criticality have a small but finite growth rate, the
densities $m_k$ have to be measured with respect to the actual lowest
level of the interface, i.e. in a co-moving frame.  The slopes of the
curves in Fig.~\ref{FigFiniteSizeMC} give an estimate of
$x_k/\nu_\perp$ (and therewith $x_k$, see right column of Table~I).
In addition, the finite growth rate measured in this type of
simulation is expected to scale as $v \sim N^{-y/\nu_\perp}$ which
allows an estimate of the exponent $y$ in Eq.~(\ref{velscal}). Our
results are $y=1.71(5)$ for the unrestricted and $y=1.76(10)$ for the
restricted model (see Fig.~\ref{FigFiniteSizeMC}) which is in
agreement with our scaling prediction $y= \nu_\parallel \simeq 1.734$ in
Eq.~(\ref{velscal}).

\begin{figure}
\epsfxsize=90mm         
\epsffile{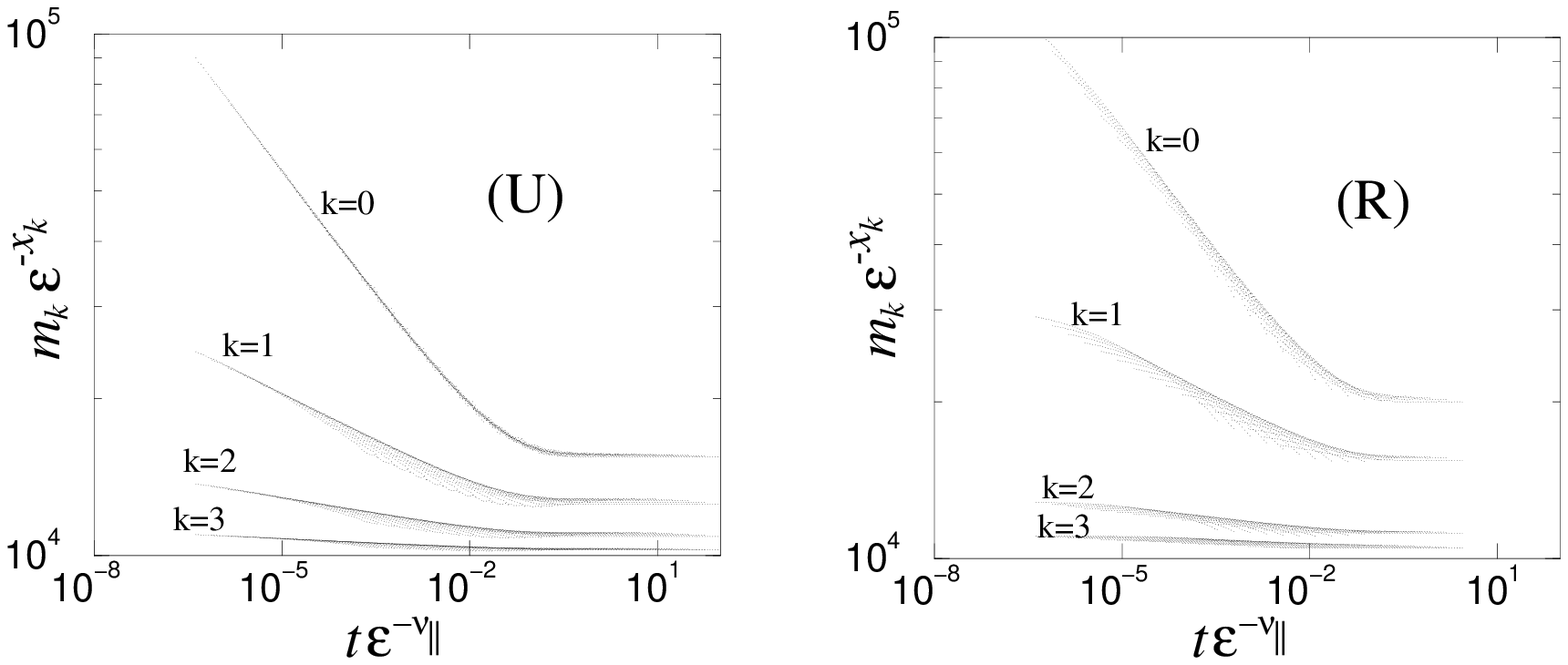}
\caption{
Finite time static simulations: Data collapse 
for the scaled integrated densities
$m_k$ as a function of time measured in the smooth phase for various
values of $\epsilon$ in the case of (U) the unrestricted and (R) the
restricted growth model.}
\label{FigStaticTimeMC}
\end{figure}

As a final check of scaling we can also obtain a finite time collapse
using the short time data from static simulations.  In the smooth
phase the expected scaling is $m_k(t) \sim |\epsilon|^{x_k} f_k(t
|\epsilon|^{\nu_{||}})$, where $f_k(s) \rightarrow const$ for $s
\rightarrow \infty$ and $f_k(s) \sim s^{-x_k/\nu_{||}}$ for $s
\rightarrow 0$. Fig.~\ref{FigStaticTimeMC} shows the scaling functions
$f_k(s)$ for various values of $\epsilon$ ranging from $-10^{-4}$ to
$-0.07$ measured up to $10^4$ time steps.  The long time stationary
values of the densities $m_k$ in static simulations (\ref{Method2})
correspond to the saturation levels of the different curves.
%
%
%

Since the three different methods lead to the same results for $x_k$
within numerical errors, we conclude that the scaling exponents
$\nu_\perp^{(k)}$ and $\nu_{||}^{(k)}$ are indeed identical on all
levels and equal to the DP exponents $\nu_\perp$ and $\nu_{||}$.  We
obtain the DP exponent $x_0=\beta$, as expected at the bottom layer,
while $x_1\simeq 0.12,x_2 \simeq 0.04,x_3 \simeq 0.015,\ldots$.  These
latter exponents, although showing the same trend, are
clearly different  numerically from those obtained using the 
heuristic picture of subsection~\ref{NumSubsectionA}.  
It is not clear whether $\{ x_k \}$ for $k > 0$ are
related to the DP exponents or whether they are independent
exponents.  The fact that we obtain the same exponents in the
restricted and the unrestricted model suggests that both variants --
at least with respect to the first few layers --
belong to the same universality class.

\subsection{Scaling  of the Interface Width}
%
%
%
\begin{figure} 
\epsfxsize=85mm        
\centerline{\epsffile{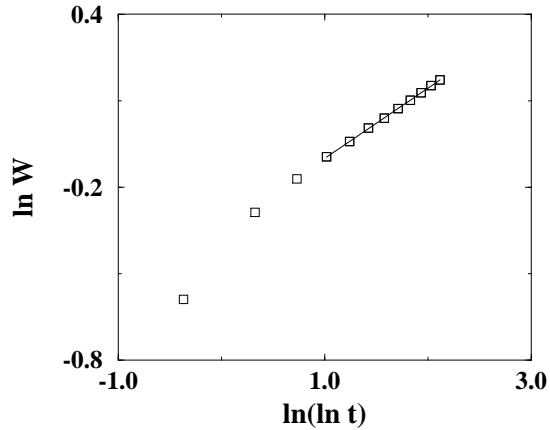}}
\caption{ 
Width at criticality as a function of time in the unrestricted model.
The  graph is for $N=2048$ and shows the growth of the width starting
from a flat interface averaged over 2000 runs. The straight line is a
best fit through the long time points ($2^6$---$2^{13}$ MCS) and has
a slope of 0.24.} \label{WidthCritU} \end{figure}
%
%

%
%
%
\begin{figure}
\epsfxsize=80mm         
\centerline{\epsffile{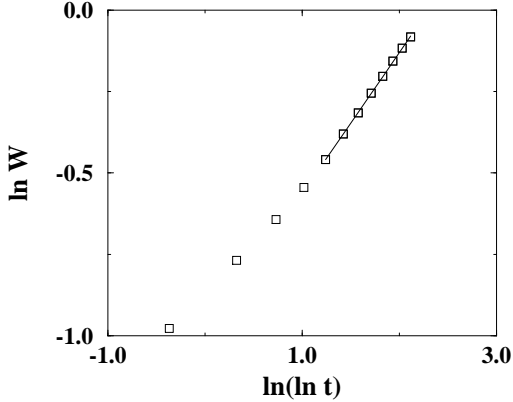}} 
\caption{  Width at criticality as a
function of time in the restricted model. The graph is as for
Fig.~\ref{WidthCritU} and the system size is $N=2048$.  The best fit
through the longer time points ($2^6$--- $2^{13}$ MCS)  has a slope of
0.43.}
\label{WidthCritR}
\end{figure}

In this subsection we investigate the scaling properties
of the interface width, defined by (\ref{width}),
at criticality and as  the rough phase is entered. The latter
measurements lead us to conclude that a complicated
scaling behavior prevails.

First recall that in the smooth phase the interface
explores only a finite number of levels and the width is finite. In
the rough phase the width is expected to diverge according to $W \sim
N^{1/2}$, the behavior generic to moving interfaces in one dimension
\cite{KPZ}.  

A naive scaling picture would suggest that the interface width may be
written as $W \sim |\epsilon| ^{-\eta} f(N/{\xi_\perp}) \sim
|\epsilon| ^{-\eta} f(N |\epsilon |^{\nu_\perp})$, where $\eta$ is
some critical exponent.  We refer to this picture as naive because
implicit is the simple scaling assumption of a single length
scale in the problem, a point we shall question in Sec. IV D.
Within this naive picture one chooses a scaling function $f$ so as to
obtain the expected $\sqrt N$ behavior in the rough phase. This
leads to the
following small $\epsilon$, large $N$ asymptotic limits:
\begin{equation}
\label{NaiveScalingPicture}
\begin{array}{lcc}
W \sim |\epsilon|^{-\eta}       & \mbox{for} &  \epsilon < 0\\
W \sim N^{\eta / \nu_{\perp}}   & \mbox{for} &  \epsilon = 0\\
W \sim N^{1/2} \, \epsilon ^{\nu_{\perp}/(2 -\eta)} 
                                & \mbox{for} &  \epsilon >0
\end{array}
\end{equation}
We now proceed to examine the actual behavior of the width at criticality.
In Figs.~\ref{WidthCritU}--\ref{WidthCritR},
width against time is plotted for simulations run at criticality. For
both unrestricted and restricted models  a long time
scaling behavior emerges before the width 
saturates due to the finite length of the system. For times
shorter than the saturation time, the width at criticality, 
$W_c$, grows as
\begin{equation}
W_{\rm c} \sim \left( \ln t \right)^\gamma. 
\label{gamma}
\end{equation}
Now since the saturation time
is expected to scale as $N^z$, where $z$ is the dynamic exponent,
one deduces that the saturation scaling of the width is
\begin{equation}
W_{\rm c} \sim \left( \ln N \right)^\gamma\,,
\label{gammasat}
\end{equation}
where $\gamma$ is given by $\gamma \simeq 0.24 \approx 1/4$ for the
unrestricted model and $\gamma \simeq 0.43$ for the restricted model.
These numerical results suggest that the critical exponent $\eta$ in
Eq.~(\ref{NaiveScalingPicture}) is in fact equal to zero. For the
restricted model the scaling of the width at criticality is similar to
that of an {\it unrestricted} polynuclear growth model~\cite{KW} where
$\gamma=1/2$.  The relationship between the present models and PNG
models will be discussed in Sec.~\ref{PNGSection}. However, for the
unrestricted model the value of $\gamma$ is clearly distinct from
$1/2$ which shows that the critical width could display non-universal
behavior. On the other hand $\gamma$ could be restricted to a finite
number of possible values. It should also be noted that in preliminary
simulations the value of $\gamma$ was erroneously assumed \cite{Us}.

\begin{figure}
\epsfxsize=90mm         
\epsffile{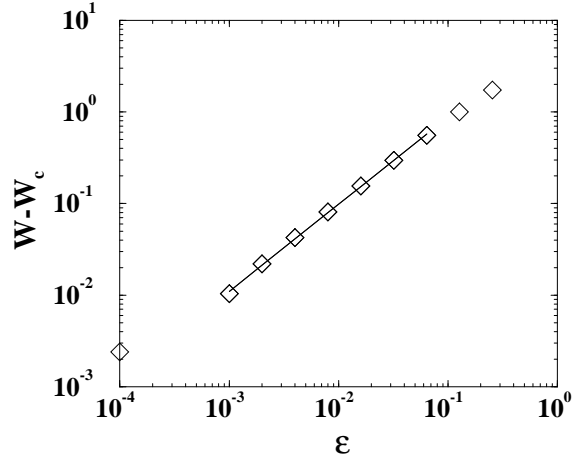}
\caption{ 
Double logarithmic plot of width in the moving phase as a
function of $\epsilon$.  The system size was $N=512$ and each point is
an average over 2000 simulations. The simulations were started from a
flat interface and allowed to saturate over $2^{13}$ MCS. The width
was then averaged over another $2^{13}$ MCS.
The straight line corresponds to $\chi =0.95$.
Simulations of the
restricted model yielded a similar behavior and estimated exponent
value $\chi =0.92$}
\label{WidthGrow}
\end{figure}

We now examine how the saturation width diverges as the growth rate is
increased and the interface enters the moving phase. To do this it is
convenient to subtract out the critical width and measure $\Delta
W(\epsilon)= W(\epsilon) - W_{\rm c}$.  Since $W_c$ is negligibly
small as compared with $W(\epsilon)$ for $\epsilon > 0$ one expects
$W$ and $\Delta W$ to have the same asymptotic behavior. In
Fig.~\ref{WidthGrow} it is seen that the scaling behavior is
\begin{equation}
\Delta W(\epsilon) \sim \epsilon^\chi\,,
\label{WidthEps}
\end{equation}
where $\chi=0.95(5)$ for the unrestricted model.
%
%
%

This result for $\chi$ disagrees with the naive scaling
picture~(\ref{NaiveScalingPicture}) which suggests that the width
should diverge with the exponent $\nu_{\perp}/2 = 0.55$ (given that
$\eta$ was found to be equal to zero).  We are left to conclude that a
more complex behavior than simple scaling takes place.

\subsection{Length Scales in the Moving Phase}
%
%
In the previous subsection we saw that a simple scaling argument
involving only the DP scaling length does not describe correctly the
numerical results. In this subsection the aim is to identify possible
additional length scales and indicate that complex critical behavior
may be present.  Therefore the subsection is by nature speculative.

We investigate scaling lengths of the first few layers in the moving
phase. We consider the correlation functions $\langle m_k (i) m_k(i+r)
\rangle$ of the integrated densities $m_k$ (\ref{IntegratedDensities})
between sites $i$ and $i+r$.

Let us first consider the bottom layer density $m_0=n_0$.
In the {\em smooth} phase it has already  been noted the
dynamics of the bottom layer (of the unrestricted model) is exactly
that of DP in the active phase.
However, in the {\em moving} phase there
is a subtle difference between the present model
and DP in the absorbing phase. This difference stems from the
fact that in the present model there is no absorbing state.
Instead, whenever a layer is completed the next highest layer
becomes the bottom layer. We effectively move
in a frame co-moving with the lowest uncompleted layer
and relabel the layers appropriately. 
%
%
%
\begin{figure}
\epsfxsize=90mm         
\epsffile{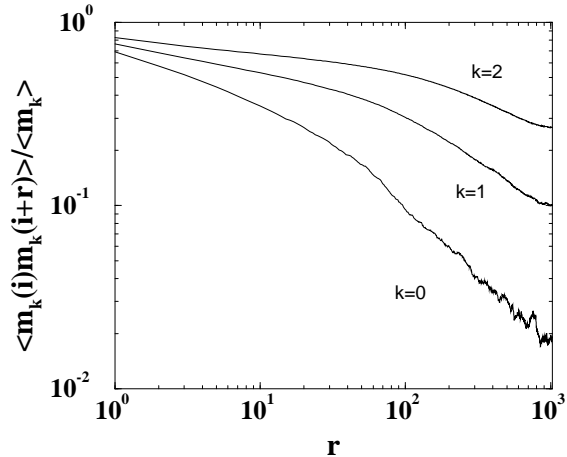}
\caption{ 
Correlation functions  
$\langle m_k (i) m_k(i+r) \rangle /\langle m_k \rangle$ 
in the moving phase in the unrestricted 
model. The system parameters are $N=2048$, 
$\epsilon=0.02$ and the results are an average over 1000
simulations each run for $2^{16}$  MCS to equilibrate. }
\label{Correlation}
\end{figure}

In order to show that this produces a non-trivial effect we plot
$\langle m_k (i) m_k(i+r) \rangle/\langle m_k \rangle$ in
Fig.~\ref{Correlation}. First recall that in the DP absorbing phase
near criticality one expects to see a power law decay $\langle n_0 (i)
n_0(i+r) \rangle /\langle n_0 \rangle \sim \langle m_0 (i) m_0(i+r)
\rangle/\langle m_0 \rangle \sim r^{-\beta/\nu_\perp}$ of the
correlation function up to a scaling length which diverges as
$\epsilon^{-\nu_\perp}$. At lengths longer than the scaling length the
correlation function should decay exponentially with $r$. In
Fig.~\ref{Correlation} for $k=0$ we see an initial power law decay
with power given by $-0.27$, then a crossover at $r \sim \xi_\perp$ to
a new power law decay with power $\simeq -0.76$ (rather than to an
exponential decay as would be the case with usual
DP scaling).  We checked for different
system sizes and values of $\epsilon$ that this qualitative behavior
(crossover to a new power law at long distance) was reproduced. Also
on Fig.~\ref{Correlation} correlation functions for the integrated
densities at higher levels are plotted and again one sees crossovers
between two power laws. The two powers appear distinct for each level
$k$. For $k=1$ the two powers are 0.15,0.56 and and for $k=2$ the two
powers are 0.09,0.36. It appears that the crossovers occur at a length
dependent on $k$ although it is difficult to quantify this. If the
crossover lengths were $k$ dependent then it would imply different
length scales existing on different levels

In order to give a heuristic explanation for the above behavior let us
consider the density at the bottom layer $m_0 =n_0$.  A picture that
could explain the crossover phenomena in  Fig.~\ref{Correlation} is
that the sites at the bottom level are grouped into clusters. Each
cluster displays the scaling behavior of an active DP cluster and
therefore is of typical size $\epsilon^{-\nu_\perp}$ and within a
cluster the correlation function decays as $\sim r^{-\beta/\nu_\perp}$.
Thus the correlations within clusters generate the first power law
decay.  However, one also has correlations between clusters. Thus
within the cluster picture the second power law measures the decay of
correlations between clusters.

A similar cluster picture could hold for the first and second layers
$k=1,2$. Within the picture, the sites at the first level, for example,
are distributed in clusters. The first power law in the $k=1$ curve of
Fig.~\ref{Correlation}  measures the correlation within a cluster and
the second measures correlations between clusters. The fact that the
crossover appears to occur at a different  distance $r$  than for the
bottom layer indicates that the clusters at the first level are larger
than those at the bottom level. Therefore there is more than one length
scale in the problem.

This picture is far from being complete or verified unambiguously   
and many questions remain open. For example it is not clear whether the
second power laws in Fig.~\ref{Correlation} continue indefinitely or
are cut off at some larger length. The numerical value of $\chi$ is
also not explained.

\subsection{Summary of the Scaling Picture}
%
For the sake of clarity we summarize the scaling results of this
section and the partial picture of scaling. First we have the smooth
phase with exponents $x_k$ associated with the density at each level. 
$x_0$ is given by the DP exponent $\beta$ whereas $x_k$ for $k \geq 1$
appear to be  non-trivial, in the sense that they are not simply
related to DP exponents. The simple approximation of 
Sec.~\ref{NumSubsectionA} gives the qualitative trend but is ruled out
quantitatively by the numerics.  As the roughening transition is
approached the DP scaling length $\xi_\perp$ and scaling time
$\xi_\parallel$ hold at all levels. This implies that approaching the
transition the dynamic exponent is $z = \nu_\parallel/\nu_\perp$.

At the transition the interface has logarithmically diverging width of
the form (\ref{gammasat}). However the  value of $\gamma$ appears to
depend on which version of the model one simulates. This could either
point to $\gamma$ being non-universal or that it can take one of a
finite number of values.

Above the transition the velocity grows with the DP exponent
$\nu_\parallel$. This reflects presence of the DP scaling length and
time at the bottom level. However by measuring the growth of the
interface width we have ruled out a simple scaling picture involving
only the DP scaling length. We have provided evidence to suggest that
there may be longer scaling lengths which characterize the size of 
clusters of sites at higher levels. This picture remains to be fully
investigated. Likewise the question of  scaling times for different
levels in  the rough phase has not been fully addressed. (In the rough
phase the dynamic exponent should take the KPZ value $z=3/2$).

It is interesting to note that in a very recent renormalization group
analysis of the field theory introduced in
Sec.~\ref{FieldTheorySection}, multicritical behavior was identified
\cite{Taeuber}. This could be consistent with the complicated scaling
behavior we have observed.


\section{Spontaneous symmetry breaking}
\label{SSBSection}
%
%
The growth models defined in Sec.~\ref{DefinitionSection}
may be viewed as examples of spontaneous symmetry breaking (SSB)
in a 1d system with periodic boundary conditions and a non-conserved
order parameter. As will be shown, the models display a robust local
mechanism for eliminating islands of the minority phase generated by 
fluctuations.

The symmetry of the growth models, apart from spatial translation
and reflection invariance, is a (discrete) translational invariance in
the heights ($Z_\infty$). In the smooth phase this symmetry is
broken since the system spontaneously selects one of the heights 
as a reference level which then serves as bottom layer for local 
fluctuations of the smooth interface.

\vspace{3mm}
\subsection{Order parameters.}
%
In order to quantify the symmetry breaking, we define a
magnetization-like 
order parameter (valid for both the restricted and the unrestricted
models):
\begin{equation}
\label{OrderM1}
M_1 = \frac{1}{N} \sum_{j=1}^N (-1)^{h_j}\,.
\end{equation}
This order parameter is clearly not conserved by the 
dynamical rules of the models. In the smooth phase $q<q_c$, it has
a nonvanishing expectation value $\langle M_1 \rangle \neq 0$ in the
thermodynamic limit where $\langle \ldots \rangle$ denotes
thermal average. On the other hand, in the rough phase
the interface explores many height levels therefore
the contribution to the magnetization at different
sites are uncorrelated over long distances
and $\langle M_1 \rangle=0$. Near the phase transition
we expect the order parameter to vanish as
\begin{equation}
\langle M_1 \rangle \sim |\epsilon|^{\theta_1}\,,
\end{equation}
where $\theta_1$ is the associated critical exponent.
%
%
%
%
\begin{figure}
\epsfxsize=85mm         
\epsffile{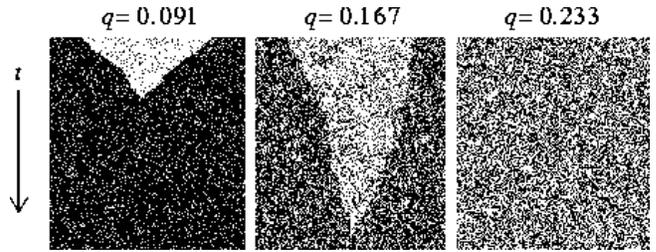} \vspace{2mm}
\caption{ 
MC simulation of the restricted model for a systems of size
600 at different values of $q$. Each configuration is a row of pixels,
with sites at even (odd) heights represented by black (white) pixels,
visualizing the order parameter $M_1$.  Configurations at intervals of
7 moves per site are shown up to 2100 sweeps. Initially, a large
island of size 400 is introduced. For $q < q_c$, the island shrinks
and disappears, illustrating the mechanism that insures long-range
order in the smooth phase. The typical time it takes for the island to
disappear depends on $q$, it increases and finally diverges when $q
\rightarrow q_c$. Similarly the magnitude of the order parameter $M_1$
(visualized by the greyscale contrast between the two phases)
decreases. At criticality the order parameter is zero so that the
island is not visible any more.  }
\label{FigIslandsShrink}
\end{figure}

The order parameter and the SSB mechanism are illustrated in
Fig.~\ref{FigIslandsShrink}. 
Here the heights are represented by alternating black and
white coloring (c.f. Fig.~\ref{Deffig})
and therefore the average greyscale in the figure
is related to the magnitude of $M_1$. Also
it is shown how  a  large island of one phase shrinks
when introduced into a system dominated by the other phase,
thus ensuring the stability of 
the smooth phase. This behavior is typical of
islands of all sizes, except for those extending over the whole system.

\begin{figure}
\epsfxsize=90mm         
\epsffile{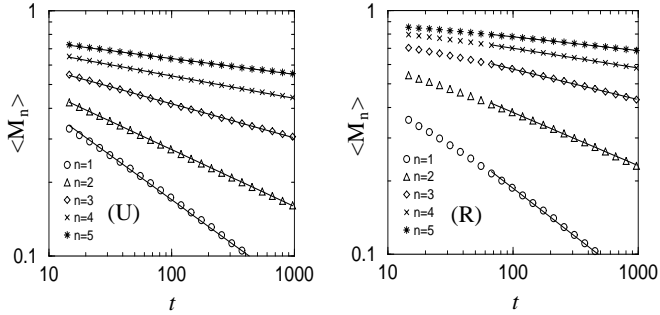}
\caption{ Measurement of the order parameters $\langle M_n \rangle$ in
dynamic MC simulations for (U) the unrestricted and (R) the restricted
variant of the growth model.  }
\label{FigOrderParameters}
\end{figure}

Since the symmetry of the model in the heights is $Z_\infty$, we can 
define a  family of order parameters which generalize $M_1$:
\begin{equation}
\label{OrderParameter}
M_n = \left|
\frac{1}{N} \sum_{j=1}^N \exp\Bigl(\frac{2 \pi i \, h_j}{n+1}\Bigr)
\right| \,.
\end{equation}
These order parameters have the same qualitative behavior as $M_1$
and can be understood as discrete Fourier transforms of the height
probability distribution. It turns out that
each order parameter is characterized 
by a different critical exponent:
\begin{equation}
\langle M_n \rangle \sim |\epsilon|^{\theta_n}\,.
\end{equation}
As in the case of the density exponents $x_k$, we determined the
exponents $\theta_n$ numerically by static, finite-size and dynamic
MC simulations. The most precise data are obtained from
dynamic simulations (see Fig.~\ref{FigOrderParameters}).
The numerical estimates for $\theta_k$ are summarized in
Table~II. It seems that these exponents are non-trivial
in the sense that they cannot be related in a simple manner
by scaling relations to the known DP exponents.

\begin{table}
\begin{tabular}{|c||c|c|c|c|c|}
        & $\theta_1$ & $\theta_2$ & $\theta_3$ & $\theta_4$ & $\theta_5$
        \\\hline
unrestricted: & $0.64(3)$ & $0.40(2)$ & $0.24(2)$ & $0.15(1)$ & 0.11(1)
\\
restricted:   & $0.66(6)$ & $0.37(4)$ & $0.21(4)$ & $0.14(3)$ & 0.10(2)
\end{tabular}
\vspace{2mm}
\caption{ Numerical estimates for the order parameter exponents
$\theta_k$ obtained from dynamic MC simulations.}
\end{table}
%
%
%
%
%
\vspace{3mm}
\subsection{External ordering fields.}
\begin{figure}
\epsfxsize=60mm         
\begin{center}~
\epsffile{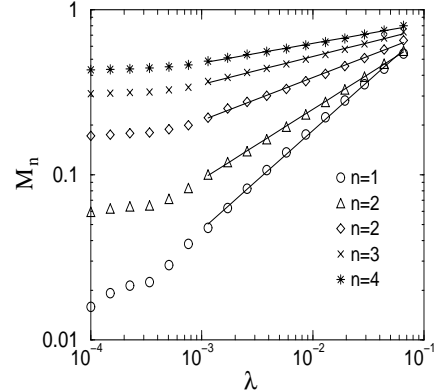}
\end{center}
\vspace{-4mm}
\caption{ Response of the order parameter $M_1$ to an external
ordering field at the critical point in the unrestricted model with
$2000$ sites. Measuring the slope of the line in the
double-logarithmic plot over two decades
gives estimates for the response exponents $\kappa_n$ (see text).
The saturation for very low values of $\lambda$ seems to be a
finite-size effect.}
\label{FigExternalField}
\end{figure}

%
For each order parameter $M_n$ one can define a conjugate
ordering field that favors states where the order parameter is positive.
This ordering field can be introduced by a periodic modulation 
of the growth rate, {\it i.e.} we replace the uniform growth rate $q$
by a height-dependent growth rate
\begin{equation}
\label{Conjugate}
q \rightarrow q(h_i) = q-\lambda \cos 
\Bigl( \frac{2 \pi \, h_i}{n+1} \Bigr)\,,
\end{equation}
where $\lambda$ is the magnitude of the ordering field.
For example, for $n=1$ and $\lambda >0$ 
the growth on the bottom layer
and other  even layers is penalised whereas growth on
odd layers is favored. At criticality,
the response to this external field 
is expected to obey a power-law behavior
\begin{equation}
M_n \sim \lambda^{\kappa_n}\,.
\end{equation}
We measured the exponents $\kappa_n$ in static MC simulation 
at criticality (see Fig.~\ref{FigExternalField}). Varying $\lambda$
over two decades from $10^{-3}$ to $10^{-1}$ we obtain the
estimates $\kappa_1=0.60(4)$,  $\kappa_1=0.42(3)$,  $\kappa_1=0.26(3)$, 
$\kappa_1=0.17(3)$, and  $\kappa_1=0.12(3)$. Comparision with the
results in Table~II suggest that $\kappa_n=\theta_n$. 

%
%
\subsection{Spontaneous breaking of continuous symmetries.}
%
So far we have shown that the growth models discussed in this paper
exhibit spontaneous breaking of a {\em discrete}
symmetry in one dimension. It is therefore of
interest to ask whether a {\em continuous symmetry} can also be broken
in one-dimensional nonequilibrium models. Continuous symmetries, where
the order parameter can take a continuum of values, seem to be harder
to break than discrete symmetries. Consider for example the equilibrium
case: discrete symmetries can be broken above one dimension; continuous
symmetries, however, can be broken only above two dimensions,
although weaker vortex ordering transitions are
possible in 2d (Kosterlitz-Thouless transitions~\cite{KT}).

Here we present a version of the growth model with a continuous
symmetry which is broken spontaneously in the smooth phase. The only
difference between this version and the unrestricted model described
above is that the height increment  at a growth move is a continuous
rather than a  discrete number.

The model is defined on a 1d lattice with periodic
boundary conditions and continuous (real) height variables $h_i$ at
sites $i=1\ldots N$. The interface evolves by choosing a site $i$ at
random and carrying out one of the processes
\begin{eqnarray}
h_i \rightarrow h_i+\zeta && \quad \quad \mbox{with probability $q$}
\nonumber \\
h_i \rightarrow \min(h_i,h_{i+1})
&& \quad \quad \mbox{with probability $(1-q)/2$} 
\label{contsymdyn}
\\ \nonumber
h_i \rightarrow \min(h_i,h_{i-1})
&& \quad \quad \mbox{with probability $(1-q)/2$} 
\end{eqnarray}
where $\zeta$ is a positive real random number selected from a flat
distribution between $0$ and $1$. The symmetry of this model (apart
from spatial translations and reflections) is continuous translational
invariance in the heights, i.e. overall shifts of the interface heights
by any amount. The symmetry breaking corresponds to the fact that in
the smooth phase the interface selects a given height which is a
real number, and remains pinned to that height for a time that grows
exponentially with the system size.

Starting from a flat interface at height zero, the dynamics taking place
at the lowest level in the continuous model are exactly the same as in
the
unrestricted version of the discrete model, the simple reason being that
in both cases each height level is decoupled from the higher ones.
This means that both models have the same critical point
$q_c=0.23267(3)$. Moreover, they have the same occupation of the
lowest level which is characterized by $n_0=(q_c-q)^{x_0}$ where
$x_0=\beta \simeq 0.277$ is the DP density exponent. 

\begin{figure}
\epsfxsize=90mm         
\epsffile{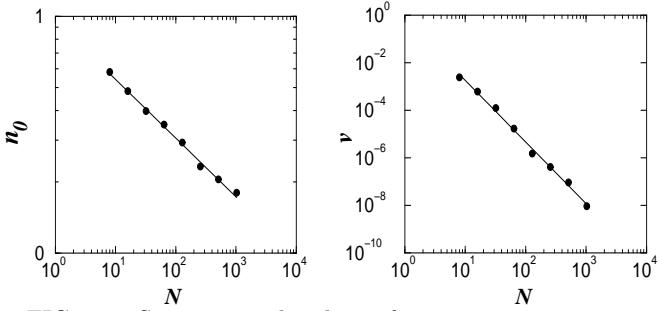}
\caption{
Spontaneous breaking of a continuous symmetry: The graphs show the
occupation of the bottom layer $n_0$ and the growth velocity~$v$
in a system of size $N$ at criticality. From the slopes we obtain
the exponents $x_0/\nu_\perp=0.245(10)$ and $y/\nu_\perp=2.59(7)$.
}
\label{FigContScaling}
\end{figure}

Above the critical point, the interface is rough and has a finite
growth velocity. Simulations show that the roughness exponent 
characterizing the interface at this phase is consistent with the
KPZ exponents~\cite{KPZ}, within 
numerical accuracy. However, an interesting
difference from the discrete model occurs in the growth velocity~$v$.
As in the previous case, it is related to the inverse of the island
lifetime. However in the present case, the step size by which the interface
grows is not $1$, but rather a real random number between $0$ and $1$.
 Each completed layer corresponds to the growth of the interface by the
smallest surviving step. Now, from our results on the discrete
models we expect  that at criticality the width behaves
as some power of $\ln N$. 
Since the number of steps in a finite system is of the order of $N$,
we expect the smallest of them to be of order $1/N$. At the critical
point, therefore, we expect the following finite size scaling for the
velocity
\begin{equation}
v \sim N^{-\nu_{||}/\nu_\perp-1}
\end{equation}
which corresponds to the scaling
\begin{equation}
v \sim (q-q_c)^y\,, \quad\quad\quad y=\nu_{||}+\nu_\perp \simeq 2.83\,.
\end{equation}
This has to be compared with $y=\nu_{||} \simeq 1.73$ in the discrete
model. These exponents are in good agreement with the values measured
in MC simulations, as shown in Fig.~\ref{FigContScaling}.
%
%
%
It would be interesting to study order parameters $M_\omega$
which are continuous generalizations of $M_n$ in 
Eq.~(\ref{OrderParameter}):
\begin{equation}
\label{ContOrderPar}
M_\omega = \left|
\frac{1}{N} \sum_{j=1}^N \exp(2 \pi i \omega \, h_j)
\right| \,.
\end{equation}
As before, we expect a power law behavior
$\langle M_\omega \rangle \sim |\epsilon|^{\theta(\omega)}$
where $\theta(\omega)$ depends continuously on $\omega$.
%
%
%
\section{Mean field approximation}
\label{MeanFieldSection}

In this section we derive the mean field equations corresponding to the
model (\ref{dyn1})--(\ref{dyn3}) and study the
resulting steady state distribution. We consider the unrestricted SOS
model since the equations for this case are
somewhat simpler. However, both the restricted and the unrestricted
models
are expected to
exhibit the same qualitative dynamical behavior.

To derive the mean field equations
we introduce at each site $i$ a set of variables $\psi_k (i)$,
$k=0,...,\infty$. Here $\psi_k (i)$ is equal
to 1 if the interface at site $i$ is at height $k$ and it is equal to 0
otherwise. Let $\langle\psi_k (i)\rangle$ be the
average of $\psi_k (i)$ over all realizations of the dynamical equations
starting with the same initial
configuration. Let us first consider the occupation of the zeroth level.
The adsorption and desorption
processes defined in Eqs.~(\ref{dyn1})--(\ref{dyn3})
result in the following equation for $\langle\psi_0 (i)\rangle$:
\begin{eqnarray}
\label{PsiZero}
\Lambda {\langle\partial  \psi _0 (i)\rangle  
\over  \partial t } =  & - & q \langle\psi _0
(i) \rangle \\ \nonumber &+& {1 \over 2} (1-q) \langle\psi _0 (i)
(1-\psi _0 (i-1) \rangle \\
\nonumber & + & {1\over  2} (1-q) \langle
\psi _0 (i) ( 1- \psi _0 (i+1)\rangle\,,
\end{eqnarray}
where $\Lambda$ is a time constant,
which, for simplicity, may be taken as $1-q$.
Within the mean field approximation, 
one replaces the correlation function
$\langle\psi_0 (i) \psi_0 (j)\rangle$ by the
product $\langle\psi_0 (i)\rangle \langle\psi_0 (j)\rangle$. 
For the ring geometry considered in
this work $\langle\psi_0 (i)\rangle$ is
independent of the site index $i$. Denoting 
$\langle\psi_0 (i)\rangle$ by $\rho_0$ we
obtain the following dynamical equation for $\rho_0$
\begin{equation}
\label{RhoZero}
{\partial \rho _0 \over \partial t} = 
-\bar q \rho _0 + \rho _0 (1- \rho _0)\,,
\end{equation}
where $\bar q = q/(1-q)$. Similar considerations yield the 
following set of equations for
$\rho_k=\langle\psi_k (i)\rangle$ for $k \geq 1$:
\begin{equation}
\label{RhoK}
{\partial \rho_k\over \partial t}=\bar q \rho_{k-1}-\bar q \rho_k+\rho
_k
\left( 1-\sum^k_{j=0}\rho_j
\right) -\rho_k \sum ^{k-1}_{j=0}\rho _j \,.
\end{equation}
It is convenient to rewrite Eqs.~(\ref{RhoZero}),(\ref{RhoK})
in terms of the integrated density variables 
(c.f. Eq.~(\ref{IntegratedDensities}))
\begin{equation}
\label{phidef}
\phi_k = \sum^k_{j=0} \rho_j\,.
\end{equation}
Substituting (\ref{phidef})
and using $\rho_k = \phi_k-\phi_{k-1}$
brings  (\ref{RhoZero}),(\ref{RhoK}) into the form
\begin{eqnarray}
\label{PhiEquations}
{\partial \phi _0\over \partial t} & = & \epsilon \phi _0 -\phi ^2_0 \\
\nonumber {\partial \phi _k\over \partial t} & = & \epsilon \phi_k -
\phi
^2_k + (1-\epsilon ) \phi_{k-1}
~~~(k\geq 1) \,,
\end{eqnarray}
where $\epsilon = 1-\bar q$. These equations have a stationary
solution corresponding to a smooth interface for $\epsilon >0$. The
roughening transition takes place at $\epsilon = 0$. The stationary
solution for $\epsilon >0$ may be calculated by the following recursion
relation:
\begin{equation}
\label{Recursion}
\phi _k= {1\over 2} \left[ \epsilon + \sqrt{\epsilon ^2 +
4(1-\epsilon)\phi
_{k-1}}~\right]\,
\end{equation}
with $\phi_0 = \epsilon $. To leading order in $\epsilon$ 
Eq.~(\ref{Recursion}) takes the form
\begin{equation}
\label{SimpleRecursion}
\phi_k = \sqrt{\phi_{k-1}} \,.
\end{equation}
The steady state distribution corresponding 
to this recursion relation is
\begin{equation}
\label{PhiDistribution}
\phi_k = \epsilon ^{ {(1/2)^k}}  \,.
\end{equation}
Therefore the mean field values of the exponents $x_k$
defined in Eq.~(\ref{xkdef}) are given by 
\begin{equation}
x_k^{MF}=\frac{1}{2^k}.
\end{equation}
%
%
%
\begin{figure}
\epsfxsize=85mm         
\epsffile{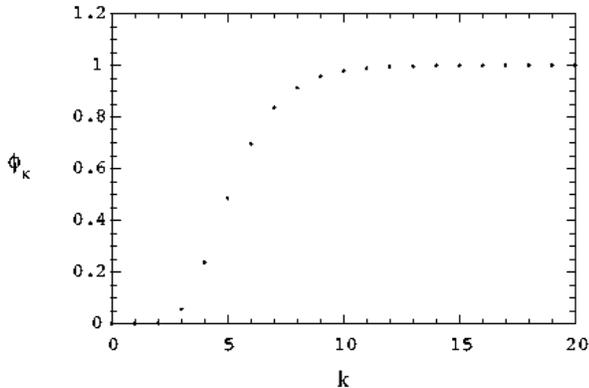} \vspace{2mm}
\caption{ Mean field approximation of the integrated height density
$\phi_k$ at level~$k$.  }
\label{FigMeanFieldA}
\end{figure}

The integrated height density $\phi_k$ is a monotonically increasing
function of $k$, varying from $\epsilon$ for $k=0$ to $1$ for
$k \rightarrow \infty$ (see Fig.~\ref{FigMeanFieldA}).
It exhibits a rapid increase
near $k \simeq k_m$ which is determined by the following equation:
\begin{equation}
\label{InflectionPhi}
\phi_{k_m -1} + \phi_{k_m +1} - 2 \phi_{k_m} =0 \,.
\end{equation}
The index $k_m$ corresponds to the interface height at which the density
$\rho_k = \phi_k - \phi_{k-1}$ is maximal. 
Using Eq.~(\ref{SimpleRecursion})
one finds that to leading order in $\epsilon$
\begin{equation}
\label{Kmax}
k_m \approx {1 \over \ln 2} \ln [ - \ln(\epsilon)] \,.
\end{equation}
It is easy to see that the width of the $\phi$ distribution remains
finite even in the limit $\epsilon \rightarrow 0$. The interval
$\Delta k$ corresponding to a change of $\phi$ from some value
$\phi_{min}$ to another, say, $\phi_{max}$ is given
to leading order in $\epsilon$ by
\begin{equation}
\label{DeltaK}
\Delta k  = {1 \over \ln 2} \ln [  \ln(\phi_{min})/ \ln(\phi_{max}) ].
\end{equation}
This interval is independent of $\epsilon$, and thus the width
of the $\phi_k$ distribution is {\em finite}. 
This feature, together with
Eq. (\ref{Kmax}) yield the following behavior for the average
height :
\begin{equation}
\label{AverageHeight}
\langle h\rangle \sim \sqrt {\langle h^2\rangle} \
\sim \ln [ - \ln ( \epsilon)] 
\end{equation}

We now turn to the magnetization-like order parameter
$M_1= \sum_j (-1)^j \rho_j$ (see Eq.~(\ref{OrderM1})). 
For small $\epsilon > 0$
Eq.~(\ref{PhiDistribution}) yields the following expression for $M_1$
\begin{equation}
\label{MeanFieldM}
M_1 = \epsilon + \sum ^{\infty}_{k=1} \epsilon^{(1/2)^k}(1-
\epsilon^{(1/2)^k}) (-1)^k\,.
\end{equation}
%
%
%
\begin{figure}
\epsfxsize=85mm         
\epsffile{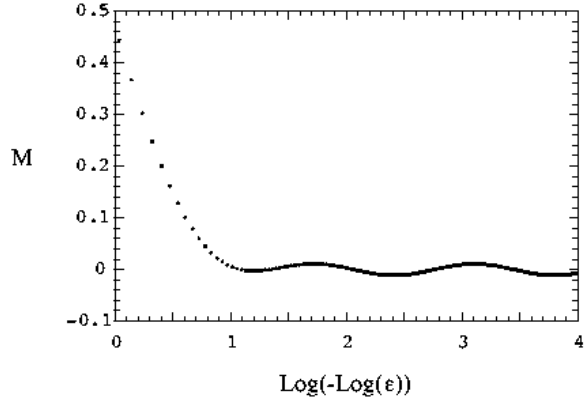} \vspace{2mm}
\caption{
Mean field approximation of the
magnetization $M_1$ as a function of $\ln[-\ln(\epsilon)]$.
}
\label{FigMeanFieldB}
\end{figure}
In Fig.~\ref{FigMeanFieldB} we plot
the magnetization as a function of $\ln (-\ln (\epsilon))$.
It is readily seen that $M_1$ does not decay to zero for small $\epsilon$.
Rather it oscillates with an amplitude which remains finite at small
$\epsilon$. This may easily be understood in the following way: the
main contribution to the sum (\ref{MeanFieldM}) comes from a finite number
of layers centered around $k= k_m$. As $\epsilon$ decreases, $k_m$
increases, and the magnetization order parameter probes different layers.
Since the contributions of the layers have alternate signs, the
magnetization becomes oscillatory. This feature is an artifact of the mean
field approximation. In this approximation fluctuations
of the position of the interface are neglected. When they are taken into
account they are expected to broaden the interface, resulting in a
diverging width at the roughening transition. This should lead to a
vanishing magnetization at the transition.
In a similar fashion, the mean field theory of KPZ interfaces does
not describe the roughening behavior.
%
%
%
\section{Field theory}
\label{FieldTheorySection}
%
%
In order to understand the universal properties observed in 
the growth models at criticality, it is useful to study the 
corresponding field theory. As will be explained below, 
such a field theory describes a hierarchy
of unidirectionally coupled DP processes. Thus it should play
a role in even more general contexts, 
namely whenever DP-like processes are coupled 
unidirectionally {\em without} feedback. 

Unlike the KPZ equation~\cite{KPZ}, the field theory we 
consider involves separate fields for each height 
level in order to incorporate the rule that atoms cannot desorb
from the middle of an island. Let us first try to guess the Langevin 
equations by adding appropriate diffusion and noise contributions
such that (a) the equation for the lowest level reproduces the ordinary
Langevin equation for directed percolation~\cite{DPLangevin}, (b)~at
a given time and position the sum over all height
densities $\sum_k\rho_k$ is equal to one, and (c)
the hierarchy of equations is translational invariant
in space and time as well as in the heights.
The simplest set of Langevin equations that meets these 
requirements reads (suppressing the arguments $x,t$)
\begin{eqnarray}
\label{FieldthBaseLangevin}
\partial_t{\rho_0} &=&
-\bar{q} \rho_0 + \rho_0(1-\rho_0)
+ \nabla^2 \rho_0 + \eta_0
\\[2mm]
\label{FieldthLangevin}
\partial_t{\rho_k} &=& \nonumber
\bar{q} \rho_{k-1} -\bar{q} \rho_k   + \rho_k
\Bigl( 1-\sum_{i=0}^{k} \rho_i \Bigr)
- \rho_k \sum_{i=0}^{k-1} \rho_i
\\ &&
+ \nabla^2 \rho_k - \nabla^2 \rho_{k-1} + \eta_k - \eta_{k-1}
~~~(k>1)\,,
\end{eqnarray}
where $\eta_k(x,t)$ are field-dependent Gaussian noise fields
with two point correlations to be specified below.
Notice that Eq.~(\ref{FieldthBaseLangevin}) is just the ordinary
Langevin equation for DP~\cite{DPLangevin}.
One can also verify that the sum over all density fields
$\sum_{k=0}^\infty \rho_k$ is a constant of motion.

As in the previous section, these equations can be simplified
by introducing integrated density fields 
$\phi_k(x,t) = \sum_{j=0}^k \rho_j(x,t)$, resulting in
\begin{equation}
\label{LangevinScaling}
\partial_t{\phi_k} =
a \phi_k - \phi_k^2 + \bar{q} \phi_{k-1}
+ \nabla^2 (\phi_k-\phi_{k-1}) + \eta_k\,,
\end{equation}
where $a=1-\bar{q}$ and $\phi_{-1}=0$. 
Interestingly, the introduction of integrated densities  also led
to a considerable improvement of the numerical results
in Sect.~\ref{NumericalSection}, suggesting that
these quantities are more natural in the context
of the present problem than the height densities $\rho_k(x,t)$.

Although the Langevin equations~(\ref{LangevinScaling})
follow quite naturally from the principles (a)-(c) stated
above, it can be dangerous to conjecture the correlations of the
noise fields $\eta_k(x,t)$. A systematic derivation of the Langevin
equations and the noise  correlations based on a bosonic operator 
formalism~\cite{SecondQuantized} will be presented in 
Ref.~\cite{Taeuber}. Dropping irrelevant operators (like $
\nabla^2 \phi_{k-1}$ in
Eq.~(\ref{LangevinScaling})) and introducing independent coefficients
for all terms it is shown that the Langevin equations are given by
\begin{eqnarray}
\label{FinalLangevin}
\partial_t{\phi_k} &=&
a_k \phi_k - \lambda_k \phi_k^2 + \bar{q}_k \phi_{k-1}
+ D_k \nabla^2 \phi_k + \eta_k
\end{eqnarray}
with noise correlations
\begin{equation}
\label{FinalNoise}
\langle \eta_k(x,t) \eta_l(x',t') \rangle =
2 \Gamma_{k,l} \, \phi_k(x,t) \, \delta(x-x') \, \delta(t-t') \,,
\end{equation}
where $k<l$. Notice that there are noise correlations
{\em between} different height levels $k,l$ which are generated in a
one-loop approximation by non-trivial mixed cubic
vertices~\cite{Taeuber}.

In Sect.~\ref{NumericalSection} we observed numerically that the
scaling exponents $\nu_\perp$ and $\nu_{||}$ are identical on all 
height levels. This observation can be verified easily within an
`improved mean field approximation' as follows.
Consider the scaling transformation
\begin{equation}
x \rightarrow \Lambda x \,, \quad
t \rightarrow \Lambda^z t \,, \quad
\phi_k \rightarrow \Lambda^{-\chi_k} \phi_k\,,
\end{equation}
where $z$ is the dynamical exponent and $\chi_k$ are the scaling
exponents of the fields $\phi_k$. 
Under rescaling, 
Eqs.~(\ref{FinalLangevin}) and (\ref{FinalNoise}) 
turn into
\begin{eqnarray}
\label{FieldthRescaledLangevin}
\partial_t{\phi_k} &=&
a_k \Lambda^z \phi_k - \lambda_k \Lambda^{z-\chi_k} \phi_k^2
\\ && \nonumber
+ \bar{q}_k \Lambda^{z+\chi_k-\chi_{k-1}}\phi_{k-1}
+ D_k \Lambda^{z-2}\nabla^2 \phi_k 
+ \eta^\prime_k\,,
\end{eqnarray}
\begin{eqnarray}
\label{FieldthRescaledNoise}
\langle \eta^\prime_k(x,t) \eta^\prime_l(x',t') \rangle &=&
2 \, \Lambda^{z+\chi_l-d} \, \Gamma_{k,l} \,\phi_{k}(x,t)
\\ && \nonumber
\times \, \delta(x-x') \, \delta(t-t') \,,
\qquad (k<l) 
\end{eqnarray}
where $d$ is the spatial dimension.
Thus an infinitesimal rescaling by $\Lambda = 1+m$
would result in a change of the coefficients
\begin{eqnarray}
\label{Infinitesimal}
\nonumber
\partial_m\, a_k &=& a_k \, z
\\ \nonumber
\partial_m\, \lambda_k &=& \lambda_k \, (z-\chi_k)
\\ \nonumber
\partial_m\, D_k &=& D_k \, (z-2)
\\ 
\partial_m\, \Gamma_{k,l} &=& \Gamma_{k,l} \, (z + \chi_l - d)
\\ \nonumber
\partial_m\, \bar{q}_k &=& \bar{q}_k \,
(z + \chi_k - \chi_{k-1})
\end{eqnarray}
At the critical dimension $d=d_c$, we expect the coefficients to be 
invariant under rescaling. As usual, the DP equation at the lowest level 
$k=0$ yields the solution $z=2$, $\chi_0=2$, and $d_c=4$. The linear
term is relevant so that the parameter $a_0$ has to be
tuned to zero (this is the mean-field critical point of DP). 
Also at higher levels $k>0$ the linear term is the most relevant contribution
wherefore $a_k=0$ is the multicritical point in mean-field.
Requiring that the nonlinearity and the coupling to the
previous level are equally relevant ($z+\chi_k-\chi_{k-1}=z-\chi_k$),
we are led to the solution $\chi_k=2^{1-k}$.  
Identifying $z = \nu_{||}/\nu_\perp$
and $\chi_k = x_k/\nu_\perp$ we obtain,
in agreement with Sec.~\ref{MeanFieldSection}, 
the mean field exponents
\begin{equation}
\nu_{||}^{MF}=1\,, \quad \quad
\nu_\perp^{MF}=\frac12\,, \quad \quad
x_k^{MF} = \frac{1}{2^{k}}\,.
\end{equation} 

The field theory (\ref{FinalLangevin})-(\ref{FinalNoise}) 
may be interpreted as a hierarchy of DP processes
which are {\em unidirectionally} coupled by the term $\bar{q}_k
\phi_{k-1}$ . 
Identifying the fields $\phi_0, \phi_1, \phi_2, \ldots$
with densities of particles $A,B,C, \ldots$,
this field theory 
corresponds to the reaction-diffusion process
\begin{eqnarray}
&& \nonumber A \leftrightarrow 2A\,, \quad A \rightarrow B
\\ && \nonumber
B \leftrightarrow 2B\,, \quad B \rightarrow C
\\ && \nonumber
C \leftrightarrow 2C\,, \quad C \rightarrow D
\\ && \nonumber \quad \ldots \quad \quad \quad \quad  \ldots
\end{eqnarray}
MC simulations indicate that this reaction-diffusion process
belongs indeed to the same universality class as the
roughening transition discussed in this paper.
We therefore expect that this field theory describes not
only the present growth models but any system in which DP 
processes are coupled in one direction.

%
%
%
%
%
\section{Relation to Polynuclear Growth Models}
\label{PNGSection}
%
%
In the previous sections we have seen that
the dynamics of the lowest layer undergoes a DP transition
that corresponds to a transition from
zero to finite velocity of the interface. For a class of models termed
Polynuclear Growth (PNG)
\cite{Rich,Goldenfeld,KW,LRWK,Toom}, which employ {\it parallel}
dynamics,
a similar scenario pertains to the growth at the highest
level. In these models the use of parallel dynamics
implies that the maximum velocity is 1 {\it i.e.}
the sites at the highest level $h=T$ grow at every time step $T$.
The sites at the highest level may be considered as active sites of
a DP process and below the transition there is a non-zero
density of such sites. Above the transition there are no
sites at the highest level and the velocity is less than~1.
This transition is lost, however, when the dynamics
are performed random-sequentially since then there
is no maximum velocity. This contrasts with the model
(\ref{dyn1})--(\ref{dyn3}) where  a phase transition is
found whether the dynamics be implemented
random-sequentially or in parallel.

In this section we  shed some light on the connection
between the present models and models of the PNG class.
In order to do this we first generalize the dynamics
of the unrestricted model (\ref{dyn1})--(\ref{dyn3}) 
to encompass both random sequential and parallel dynamics.
In a time step {\it all} sites are updated
according to the following rule:
\begin{eqnarray}
\label{dynrsp}
\lefteqn{h_i(T+1) }\nonumber \\
&=& h_{i}(T) \; \; \mbox {with prob. }\;\; 1- \Delta 
\\
 &=& h_{i}(T) +1 \; \; \mbox {with prob. }\;\; q \Delta 
\nonumber \\
&=& \mbox{min}\left[h_{i}(T),h_{i+1}(T)\right] \; \; 
\mbox {with prob. }\;\; (1-q)\Delta /2
\nonumber \\
&=& \mbox{min}\left[h_{i-1}(T),h_{i}(T)\right] \; \;
\mbox {with prob. }\;\; (1-q)\Delta /2 \nonumber
\end{eqnarray}
As $\Delta$ varies from 0 to 1, the rule (\ref{dynrsp}) 
interpolates between random sequential
dynamics and parallel dynamics: for
$\Delta  \ll 1/N$ the height of at most one site is modified at any time
step and the dynamics becomes random sequential;
for $\Delta  =1$ all sites are modified at each time step
and the dynamics is parallel.

We now make a transformation suggested to us
by J. Krug (private communication). If one defines
\begin{equation}
h_i(T)= T- g_i(T)\; ,
\label{transform}
\end{equation}
the variables $g_i(T)$ undergo the following
dynamics:
\begin{eqnarray}
\label{gdyn}
\lefteqn{g_i(T+1)}\nonumber \\
&=& g_{i}(T)+1 \; \; \mbox {with prob. }\;\; 1- \Delta \\
&=& g_{i}(T)  \; \; \mbox {with prob. }\;\; q \Delta \nonumber \\
&=& \mbox{max}\left[g_{i}(T), g_{i+1}(T)\right]+1 \; \;
\mbox {with prob. }\;\; (1-q)\Delta /2
\nonumber \\
&=& \mbox{max}\left[g_{i-1}(T), g_{i}(T)\right] +1 \; \; 
\mbox {with prob. }\;\; (1-q)\Delta /2\; . \nonumber
\end{eqnarray}
The rules (\ref{gdyn}) yield a growth model where
the maximum velocity is 1 and the DP transition
appears at the maximal height level, as occurs with
PNG models. However, it is important to note
that the dynamics (\ref{gdyn}) is always parallel in nature
since the heights of many sites are modified at each
time step. Thus both random sequential and parallel
versions of (\ref{dynrsp}) are mapped onto a parallel
rule~(\ref{gdyn}).

We now compare (\ref{gdyn}) with a specific PNG model studied
in~\cite{KW}. In that model the heights of a $1d$ interface are updated
in parallel in two sub-steps. First all up (down) steps of the
interface move deterministically to  left (right) a distance of $u$
lattice spacings. Then all heights are incremented by 1 with
probability $p$.  (For our purposes it is convenient to group the two
substeps in the reverse order to that of \cite{KW} but this does not
change the dynamics). Thus the model is unrestricted. For the case $u=1$
this dynamics may be written as a single parallel update where
\begin{eqnarray}
\label{PNGdyn}
\lefteqn{g_i(T+1)}\\
&=& \mbox{max}\left[g_{i-1}(T), g_{i}(T), g_{i+1}(T)\right]+1 \; \; 
\mbox {with prob. }\;\; p
\nonumber \\
&=& \mbox{max}\left[g_{i-1}(T), g_{i}(T), g_{i+1}(T)\right] \; \; 
\mbox {with prob. }\;\; (1-p)\;. \nonumber 
\end{eqnarray}
Clearly the two dynamics (\ref{gdyn}),(\ref{PNGdyn}) are
similar in character but distinct. 
Furthermore in~\cite{KW} it was found that
at $p_c \simeq 0.527$ the width of the interface scales
as $W \sim (\ln N)^{1/2}$ which is
distinct from  the behavior  of the unrestricted model
but  similar to that of the
{\it restricted model}  of the present paper.

In this section we have seen a subtle connection between the growth
models studied in the present paper and models similar in character to
PNG models. This was done by generalizing the dynamics of the present
model to a dynamics that interpolates between random sequential and
parallel.  Then, transforming the present model via (\ref{transform})
one obtains a model with parallel dynamics which has similarities with,
but is distinct from, the PNG models previously studied.
However, with the standard PNG model (\ref{PNGdyn}) the dynamics
is strictly parallel and it is not clear if it can be
transformed to any random sequential model. Indeed, simply
employing the inverse transform of (\ref{transform}) yields
another model with strictly parallel dynamics.
%
%
\section{Conclusions} \label{Conclusions}
%

In summary, we have studied in detail a one-dimensional stochastic
growth model with random sequential dynamics that exhibits a transition
from a smooth to a rough phase. In studying the model we have shown
that some properties may be understood directly from the scaling
behavior of DP. However, other properties and critical exponents appear
non-trivial in the sense that they seem not to be related to DP
quantities in a simple manner. Furthermore, we have introduced novel
exponents such as those relating to the magnetization like order
parameters characterizing the SSB  and the response of these order
parameters to an ordering field.

One can also think of the model in terms of a system of 
unidirectionally coupled DP processes. We have proposed a field theory 
which should describe the properties of this general class of systems.

We are left with several open questions. Firstly can the values of the
novel exponents be predicted? Secondly can the critical behavior be
described by a conventional scaling picture? In our study we have
indicated that the critical behavior may be quite complicated. This
could be consistent with multicritical behavior found in a study
\cite{Taeuber} of the field theory proposed in
Sec.~\ref{FieldTheorySection}.

We have also made a subtle connection between the present random
sequential models and models similar in character to the parallel
update PNG models such as that studied in \cite{KW}. It would be
instructive to explore this point further.

It would also be of interest to study in more detail the model
(\ref{contsymdyn}) that exhibits the breaking of a continuous
symmetry. In particular the order parameter (\ref{ContOrderPar}) has
not been fully investigated.

The growth models and their scaling behavior were investigated in
one dimension. However, it is straightforward to define the models
in higher dimensions where similar scaling behavior is expected
to hold.

Let us finally remark that it would be very interesting to search for
experimental realizations of the growth processes discussed in this
paper, in particular because of their relation to DP.
As pointed out by Grassberger~\cite{DPOpenProblems}, 
the large body of theoretical work on DP seems to be 
unbalanced by the fact that there are no experiments where DP exponents
have actually been measured. It is not yet clear whether this is
due to a lack of such experiments or to an oversimplification of
nature in DP models. The growth models, however, suggest another
category of experiments where DP exponents may be identified, namely
absorption-desorption processes where the evaporation of atoms from
the middle of completed layers is highly suppressed.\\[3mm]
\noindent
{\em Acknowledgments:}  
We thank P. Bladon, Y.Y. Goldschmidt, M. J. Howard, J. Krug, V. Rittenberg,
P. Sollich, U. C. T\"auber and N. B. Wilding for interesting
discussions. MRE is a Royal Society University Research Fellow
and thanks the Weizmann Institute for warm hospitality
during several visits when this work was in progress.
UA is supported by a Rothchild Fellowship.
This work was supported by Minerva Foundation, Munich, Germany.
%
%
\appendix
\section{An Exactly Solvable Case}
\label{SolvableCaseSection}
%
%
It is instructive to consider a case where the steady state
can be calculated exactly. This will allow us to verify that
the interface is indeed rough in the moving phase
and that if we allow some rate for desorption from
the middle of a plateau, the interface is rough independent
of whether the velocity is positive, negative or zero.

The transition rates
we consider are that of the RSOS model
presented in the introduction with the
addition of a process with rate $p$.
\begin{eqnarray}
\mbox{0\ +}    \;\; &\rightarrow &\;\; \mbox{+\ 0}  \;\;\;\mbox{with
rate}\;\;
g
\nonumber \\
\mbox{+\ 0}    \;\; &\rightarrow  &\;\; \mbox{0\ +} \;\;\;\mbox{with
rate}\;\;
1
\nonumber \\
\mbox{$-$\ 0}  \;\; &\rightarrow &\;\; \mbox{0\ $-$}\;\;\;\mbox{with
rate}\;\;
g
\nonumber \\
\mbox{0\ $-$}  \;\; &\rightarrow  &\;\; \mbox{$-$\ 0}\;\;\;\mbox{with
rate}\;\;
1
\nonumber \\
\mbox{0\ \ 0}  \;\; &\rightarrow &\;\; \mbox{+\ $-$}  \;\;\mbox{with
rate}\;\;
g
\nonumber \\
 \mbox{+\ $-$} \;\; &\rightarrow  &\;\; \mbox{0\ \ 0}  \;\;\;\mbox{with
 rate}\;\; 2
\nonumber \\
\mbox{$-$\ +}  \;\; &\rightarrow &\;\; \mbox{0\ \ 0} \;\;\;\mbox{with
rate}\;\;
g
\nonumber \\
\mbox{0\ \ 0}  \;\; &\rightarrow  &\;\; \mbox{$-$\ +} \;\;\;\mbox{with
rate}\;\; p
\end{eqnarray}
The process with rate $p$ translates to
desorption from the middle of a plateau
when the model is translated back into a growth model via
(\ref{ChargedRep}).
We shall using a technique similar to that employed recently
in~\cite{NdN} (see also~\cite{GW}) to show that if
\begin{equation}
p = 1- g/2
\label{solvcon}
\end{equation}
the steady state probabilities $P$ of a configuration may be
written in a factorized form
\begin{equation}
P(M) = Z_N ^{-1}\  2^{-M} \,,
\label{facprob}
\end{equation}
where $M$ is the number of positive charges in the configuration
and $Z_N$ is a normalization.
One also has the constraint that the only
allowed configurations have equal numbers of positive and
negative charges due to the fact that dynamics conserves the global
charge. Taking this into account the normalization $Z_N$ is given by
\begin{equation}
Z_N = \sum_{M=0}^{N/2} \frac{N!}{M! \ M! \ (N-2M)!} \, 2^{-M}\;.
\label{norm}
\end{equation}

In order to prove Eq.~(\ref{facprob}) let us define
variables $b_{+0}, b_{++}, b_{+-} \ldots$ where, for example,
$b_{+0}$ is the number of bonds $i,i+1$ where there is a positive
particle at site $i$ and a hole at site $i+1$.
Due to the fact that the global charge is zero we have
$b_{+0} + b_{++} + b_{+-} = b_{0+} + b_{++} + b_{-+}
=b_{-0} + b_{--} + b_{-+} = b_{0-} + b_{--} + b_{+-}=M$
which leads to
\begin{equation}
b_{+0} + b_{0-} + 2b_{+-} =b_{0+} + b_{-0} + 2b_{-+}
\label{conserv}
\end{equation}
If Eq.~(\ref{facprob}) is to hold in the steady state
the following equation for the balance
of probability must hold for any  configuration:
\begin{eqnarray}
\label{probbal}
\lefteqn{(b_{+0} + g b_{0+} + b_{0-} + g b_{-0} }
\nonumber \\
&&+ 2b_{+-}
 + g b_{-+} + (g +p) b_{00})P(M)  \\
=&&
(b_{0+} + g b_{+0} + b_{-0} + g b_{0-})P(M) \nonumber \\
&& + (2 + g)b_{00}P(M+1) + (g b_{+-} + p b_{-+}) P(M-1) \nonumber
\end{eqnarray}
To understand Eq.~(\ref{probbal}) note that the lhs
gives the rate of loss of probability
due to the transitions
out of the configuration and  the rhs gives
the rate of gain of probability due
to  the transitions into  the configuration.
We now divide through by (\ref{facprob})
and use (\ref{conserv}) in (\ref{probbal})
to yield
\begin{equation}
 (g +p)b_{00}  -g b_{-+}
= (1 + g /2)  b_{00} +(2p -2 ) b_{-+} \; ,
\end{equation}
which is satisfied for arbitrary $b_{00}, b_{-+}$
when (\ref{solvcon}) holds.

In order to calculate the velocity and roughness
for $N$ large one notes that the sum for the normalization
(\ref{norm}) is dominated by $M = N \ ( 1-1/\sqrt{2}) $,
therefore  $\rho$, the steady state density of positive charges
(and also that of negative charges), is
\begin{equation}
\rho = 1 - 1/\sqrt{2} + {\cal O}(1/N)\,.
\end{equation}
Here we define the velocity $v$ as the
steady state growth rate at an arbitrary site $i$.
Let $\langle c_i c_{i+1} \rangle$ be the steady state
expectation of finding a charge  $c_i$ at site $i$
and a charge $c_{i+1}$ at site $i+1$, then
\begin{eqnarray}
v&=& g\langle 0_i\ +_{i+1}\rangle
 -\langle +_i\ 0_{i+1} \rangle  \nonumber \\
&& -\langle 0_i\ -_{i+1} \rangle
  +g\langle -_i\ 0_{i+1}  \rangle \nonumber \\
&&  +(3g/2 -1)\langle  0_i\ 0_{i+1} \rangle   \nonumber \\
&& -2 \langle +_i\ -_{i+1} \rangle
  +g\langle -_i\ +_{i+1}  \rangle \,.
\end{eqnarray}
The form (\ref{facprob}) implies that correlation functions
factorize to leading order in $1/N$
(e.g. $\langle +_i\ +_{i+1} \rangle
\simeq \rho^2$) and one finds
\begin{eqnarray}
v &=& 2(g-1) \rho (1-2 \rho) \nonumber \\
&& +( 3g/2 -1 )(1-2 \rho)^2
+ (g -2) \rho^2 + {\cal O}(1/N) \nonumber \\
&\simeq & (2-\sqrt{2})(g-1) \, .
\end{eqnarray}
One can also calculate the  roughness
exactly. First we define the height at site $i$
relative to site $N$  as
\begin{equation}
h_i = \sum_{j=1}^{i} c_j\,.
\end{equation}
Clearly $\langle h_i \rangle =0$,
therefore the width $w$ is defined through
\begin{equation}
w^2 = \frac{1}{N} \sum_{i=1}^{N} \langle h_i^2 \rangle\;.
\end{equation}
A moderate calculation then yields
\begin{equation}
w^2/N = \rho/3 + {\cal O}(1/N) \simeq (1-1/\sqrt{2})/3
\label{exactrough}
\end{equation}
The result (\ref{exactrough}) implies that
when (\ref{solvcon}) holds the interface is always rough
and indeed the prefactor does not depend on $g$.

In the case $p=0$ we see from (\ref{solvcon}) that $g =2$ and we are
clearly
in the moving phase  of our original growth model.
When $g=1$ it is easy to check that we have detailed balance
so that the interface is in equilibrium and again
we expect it to be rough. Equation (\ref{exactrough}) verifies that
in the case of non-zero $p$
(and when (\ref{solvcon}) holds)  the interface is rough
whether it be moving upwards ($g>1$), downwards ($g<1$) or not at all
($g=1$).
%
%
%
%
%
%


\begin{thebibliography}{10}
\bibitem{Note}
        Here we consider a system to be out of equilibrium
        if its dynamics does not obey detailed balance.
\end{thebibliography}
\end{document}